\documentclass[ twocolumn,aps,prd,
               preprintnumbers,numbers,sort&compress,
               nofootinbib,
                            showpacs,
               colorlinks,
               linkcolor=blue,
               citecolor=blue,
               superscriptaddress]{revtex4-1}
   \newcommand{\exclude}[1]{}

\usepackage{epsfig,graphicx}
\usepackage{amsmath,amssymb,bm}
\usepackage{psfrag}
\usepackage{feynmp}
\usepackage{hyperref}

\newcommand{\beq}{\begin{equation}}
\newcommand{\eeq}{\end{equation}}
\newcommand{\be}{\begin{eqnarray}}
\newcommand{\ee}{\end{eqnarray}}

\def\dd{ \,\mathrm{d} }

\def\+{\dagger}
\def\la{\langle}
\def\ra{\rangle}
\def\<{\langle}
\def\>{\rangle}

\newcommand{\Lbar}{\Lambda_{\overline{\mathrm{QCD}}}}
\newcommand{\qcd}{{\overline{\mathrm{QCD}}}}

\newcommand{\Tr}{\mathrm{Tr}}

\begin{document}


\title {Inflation  and Gauge Field Holonomy}

\author{Andrey O. Barvinsky}
 \affiliation{Theory Department, Lebedev
Physics Institute, Leninsky Prospect 53, Moscow 119991, Russia}
\author{ Ariel R. Zhitnitsky}
 \affiliation{Department of Physics \& Astronomy, University of British Columbia, Vancouver, B.C. V6T 1Z1, Canada}


\begin{abstract}
We discuss  a novel scenario for early cosmology, when the inflationary quasi-de Sitter phase dynamically  originates from the initial quantum state represented by the microcanonical density matrix. This genuine quantum effect occurs as a result of the dynamics of the  topologically  nontrivial  sectors in a (conjectured) strongly coupled QCD-like  gauge theory  in expanding universe. The crucial element of our proposal is the presence  in our framework of a nontrivial $\mathbb{S}^1$
which plays the dual role in construction: it defines the periodic gravitational instanton (on the gravity side) and it  also defines a nontrivial gauge holonomy (on the gauge side) generating the vacuum energy.   The effect is global in nature and cannot be formulated in terms of a gradient expansion in an effective
local field theory.  We also discuss a graceful exit from holonomy inflation due to the helical instability. The number of e-folds in the holonomy inflation framework   is determined by the gauge  coupling constant at the moment of inflation, and estimated as $N_{\rm infl}\sim \alpha^{-2}(H_0)\sim 10^2$. 
 We   also comment on  the relation of our framework with the no-boundary and tunneling cosmological proposals  and their recent criticism.

 \end{abstract}

\maketitle

\section{\label{introduction}Introduction}
Inflationary scenario is widely recognized as one of the most successful candidates for the description of the early Universe leading to its observable large scale structure. Majority of effective and fundamental models of this scenario are based on the assumption that matter energy density driving the quasi-exponential expansion of the Universe during inflation stage is generated by local field-theoretical degrees of freedom, like a scalaron field in the Starobinsky $R^2$-gravity \cite{R^2} or a scalar field   inflaton $\Phi(x)$   with its potential $V(\Phi)$ in chaotic and other inflationary models \cite{inflation}, see textbook \cite{mukhanov} for a general overview.

However, it is also very possible that the generation of this type of uniformly distributed energy might not be    associated with any local propagating particles. Instead,  it might be related to  some   global characteristics (such as holonomy) or topological degrees of freedom which cannot be expressed in terms of any local fields such as inflaton $\Phi$. Examples, in particular,  include the global degree of freedom arising in the context of the recently suggested generalized unimodular gravity theory \cite{unimodular}. Another example is represented by a strongly coupled QCD-like gauge theory   when  the vacuum energy is generated by some  nontrivial topological features of the gauge systems
\cite{Zhitnitsky:2013pna,Zhitnitsky:2014aja,Zhitnitsky:2015dia}.

Here we want to apply  the ideas when  the vacuum energy is induced by the topologically nontrivial holonomy  \cite{Zhitnitsky:2013pna,Zhitnitsky:2014aja,Zhitnitsky:2015dia}
to the mechanism of inflation in the early quantum Universe driven by the thermal states \cite{Barvinsky:2006uh,Barvinsky:2006tu}. This model, which incorporates the idea of the microcanonical density matrix as the initial quantum state of the Universe \cite{why} is conceptually very attractive because of the minimum set of assumptions underlying it and, moreover, because of a mechanism restricting the cosmological ensemble to subplanckian energy domain and avoiding the   infrared catastrophe inherent in the no-boundary wavefunction \cite{noboundary}. Furthermore, this thermally   driven cosmology \cite{Barvinsky:2006uh,Barvinsky:2006tu} can serve as initial conditions for the observationally consistent models of $R^2$ and Higgs inflation,  see original paper  \cite{Higgsflation0} and the recent development \cite{Higgsflation,Higgsflation1} based on induced gravity aspects of the theory.

As we argue below our construction, which can be viewed as a synthesis of two naively unrelated ideas  \cite{Barvinsky:2006uh,Barvinsky:2006tu,why} and  \cite{Zhitnitsky:2013pna,Zhitnitsky:2014aja,Zhitnitsky:2015dia} correspondingly,  shows a number of very desirable and remarkable
features. On the gravity side \cite{Barvinsky:2006uh,Barvinsky:2006tu,why} the nontrivial element of the construction is represented by the Euclidean spacetime with a time compactified to a circle $\mathbb{S}^1$. On the gauge field theory side  \cite{Zhitnitsky:2013pna,Zhitnitsky:2014aja,Zhitnitsky:2015dia} the same  $\mathbb{S}^1$ plays a crucial role when the gauge configurations may assume a nontrivial holonomy along  $\mathbb{S}^1$.
Precisely the gauge configurations with the nontrivial  holonomy
along $\mathbb{S}^1$ may  serve as a source of vacuum energy density sustaining the inflationary scenario.
Furthermore,
as we argue below this construction provides a system  with a subplanckian energy scale such that a number of  well-known and undesirable  properties
which always accompany the conventional inflationary scenario when a system is   formulated in terms of a local field $\Phi(x)$, does not even occur in our framework.

Our presentation is organized as follows.  We begin in Sect.\ref{sec:grav-instanton} with a brief overview of the first crucial element of the proposal: the thermally  driven cosmology
when the initial state is described by the microcanonical density matrix as originally discussed in  \cite{Barvinsky:2006uh,Barvinsky:2006tu,why}.
In Sect.\ref{topology} we overview a second crucial element of our proposal  related to
fundamentally new source of the vacuum energy as suggested in
  \cite{Zhitnitsky:2013pna,Zhitnitsky:2014aja,Zhitnitsky:2015dia}.

Then in sections \ref{Inflation} and \ref{Inflation-2} we construct two different inflationary models  based on similar building principles, but   different field contexts. In both models the inflationary vacuum energy is generated by the holonomy of gauge fields.  In the first model studied in section  \ref{Inflation}  one can carry out all the computations is theoretically controllable semiclassical approximation as a result of   special selection of the matter context. The second model  studied in section  \ref{Inflation-2} is much more attractive phenomenologically, though the semiclassical approximation cannot be justified in this case.

We discuss how the  inflation ends  in our scenario (the so-called reheating epoch)   in Sect.\ref{inflation_end}.
In particular, we demonstrate that the number of $e$ folds $N_{\rm infl}$   is always very large $N_{\rm infl}\sim \alpha^{-2}(H)\sim 100$ as a result of
small gauge coupling constant $\alpha(H)\sim 0.1$ at the Hubble scale $H$. 
We also compare our holonomy inflation  with conventional description in terms of the local inflaton $\Phi$ and potential $V[\Phi]$ in section \ref{relation}.

We conclude in Section \ref{conclusion} with formulation of the basics results and profound consequences of our proposal.
We also describe in subsection \ref{test} how this new form of the topological vacuum energy can be tested  in a tabletop experiments in physical Maxwell system. We also comment in subsection \ref{no-boundary} on   differences of our framework with well known no-boundary and tunnelling proposals.
Finally,   we summarize  a number of     technical aspects relevant to our topological  inflation scenario   in Appendix \ref{review}.
In particular,  we overview the nature of the contact term in gauge theories in section \ref{contact},  the generation of the ``non-dispersive" vacuum  energy
due to the holonomy in section \ref{sec:holonomy} and its role in cosmological context in section \ref{interpretation}.

\section{\label{sec:grav-instanton}{Origin of inflation in the thermally  driven cosmology}}
Our goal here is to overview the previous results  \cite{Barvinsky:2006uh,Barvinsky:2006tu,why}
with emphasize on the periodic properties of $\mathbb{S}^1$ where gravitational instantons are defined and serve as initial conditions for the cosmological evolution of the scale factor $a(t)$. Analytical continuation to the physical Lorentzian space-time demonstrates the de Sitter like behaviour with constant $H$. This is precisely the main goal of this section.

The model of quantum initial conditions in cosmology in the form of the microcanonical density matrix was suggested in \cite{why}, where its statistical sum was built as the Euclidean quantum gravity path integral,
    \begin{eqnarray}
    &&Z=
    \!\!\int\limits_{\,\,\rm periodic}
    \!\!\!\! D[\,g_{\mu\nu},\varPhi\,]\;
    e^{-S[\,g_{\mu\nu},\varPhi\,]},         \label{Z}
    \end{eqnarray}
over the metric $g_{\mu\nu}$ and matter fields $\varPhi$ which are
periodic on the Euclidean spacetime with a time compactified to a circle $\mathbb{S}^1$. This statistical sum has a good predictive power in the Einstein theory with the primordial cosmological constant and the matter sector which mainly consists of a large number of quantum fields conformally coupled to gravity \cite{Barvinsky:2006uh,Barvinsky:2006tu}. The dominant contribution of numerous conformal modes allows one to overstep the limits of the usual semiclassical expansion, because the integration over these modes gives the quantum effective action of the conformal fields $\varGamma_{CFT}[\,g_{\mu\nu}]$ exactly calculable by the method of conformal anomaly. On the Friedmann-Robertson-Walker (FRW) background,
    \begin{eqnarray}
    ds^2=d\tau^2
    +a^2(\tau)\,d^2\Omega^{(3)},
    \end{eqnarray}
with a periodic scale factor $a(\tau)$ -- the functions of the Euclidean time belonging to the circle $\mathbb{S}^1$ \cite{Barvinsky:2006uh} -- this action is calculable by using the local conformal transformation to the static Einstein universe and the well-known trace anomaly,
    \begin{equation}
    g_{\mu\nu}\frac{\delta
    \varGamma_{CFT}}{\delta g_{\mu\nu}} =
    \frac{1}{4(4\pi)^2}g^{1/2}
    \left(\alpha \Box R +
    \beta E +
    \gamma C_{\mu\nu\alpha\beta}^2\right),     \label{anomaly}
    \end{equation}
which is a linear combination of Gauss-Bonnet $E=R_{\mu\nu\alpha\gamma}^2-4R_{\mu\nu}^2+ R^2$, Weyl tensor squared $C_{\mu\nu\alpha\beta}^2$ and $\Box R$ curvature invariants with spin dependent coefficients\footnote{ With the nonvanishing background values of matter fields there are additional contributions to the conformal anomaly like the square of the relevant Yang-Mills strength $F_{\mu\nu}^2$ or a conformal scalar field $\phi^4$, \cite{Duff}. We disregard them, because in what follows their values are assumed to be either zero or negligible compared to the contribution of the gravitational structures with large coefficients $\alpha$, $\beta$ and $\gamma$.}. The resulting $\varGamma_{CFT}[\,g_{\mu\nu}]$ turns out to be the sum of the anomaly contribution and the contribution of the static Einstein universe -- the Casimir and free energy of conformal matter fields at the temperature determined by the circumference of the compactified time dimension $\mathbb{S}^1$. This is the main calculational advantage provided by the local Weyl invariance of $\varPhi$ conformally coupled to $g_{\mu\nu}$.
\begin{figure}[h]
\centerline{\includegraphics[width=7cm]{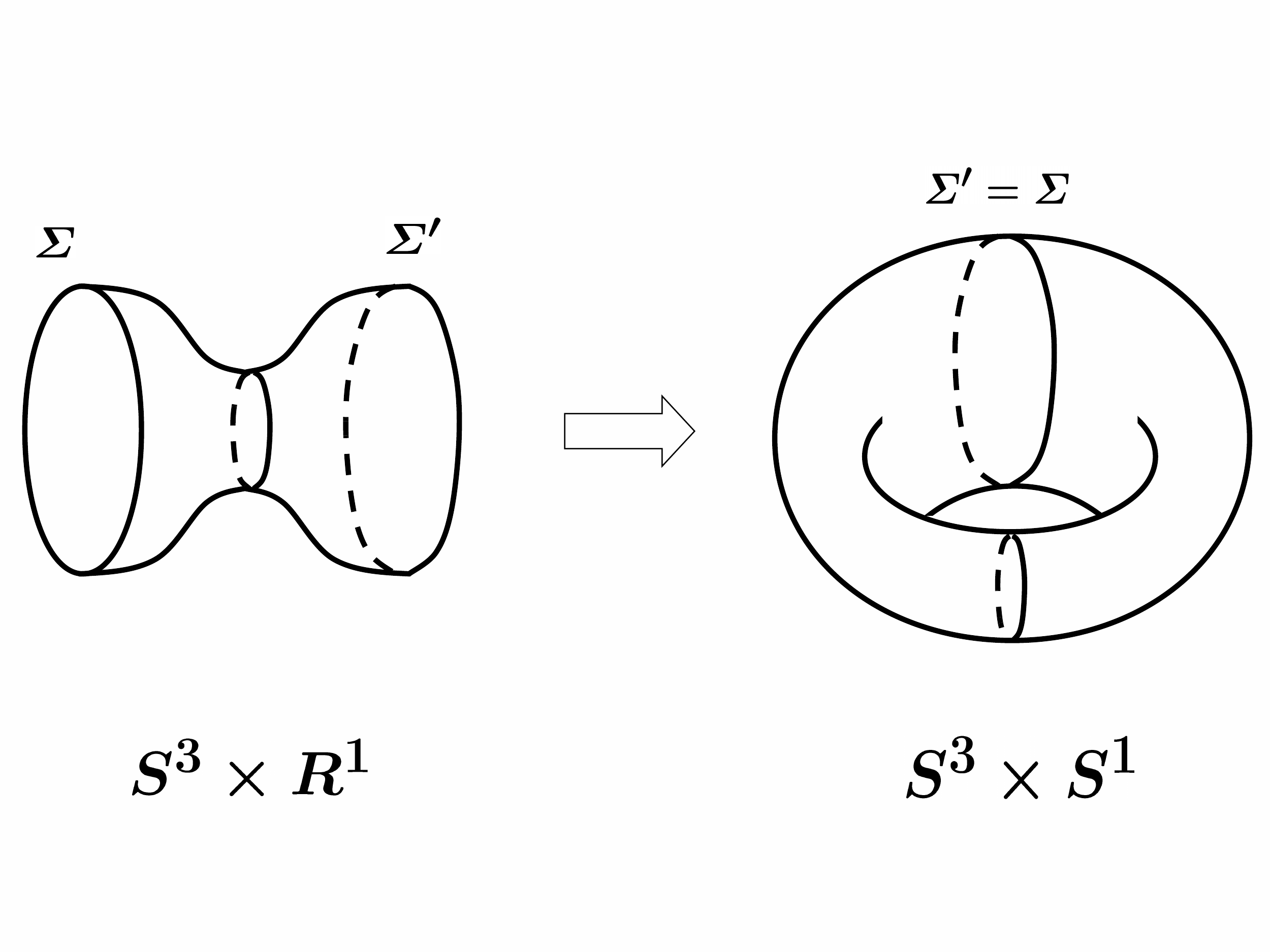}} \caption{\small
Transition from the density matrix instanton to the periodic
statistical sum instanton.
 \label{Fig1}}
\end{figure}
Solutions of equations of motion for the full effective action -- saddle points of the microcanonical statistical sum (\ref{Z}) -- are the periodic cosmological instantons of $\mathbb{S}^1\times \mathbb{S}^3$ topology (in what follows we assume spatially closed cosmology which explains spherical topology of its spatial sections). These statistical sum instantons follow by a usual tracing procedure from the two-boundary instantons of the relevant microcanonical density matrix, which is depicted on Fig.1.
In their turn, the density matrix instantons serve as initial conditions for the cosmological evolution $a_L(t)$ in the physical Lorentzian spacetime. The latter follows from $a(\tau)$ by the analytic continuation $a_L(t)=a(\tau_*+it)$ at the point of the maximum value of the Euclidean scale factor $a_+=a(\tau_*)$, as shown on Fig.2.

  This construction is described in   \cite{Barvinsky:2006uh,Barvinsky:2006tu,why} and we refer the readers to these original papers.
The only comment we would like to make here is that  the starting point of the analysis  \cite{Barvinsky:2006uh,Barvinsky:2006tu,why}  is, of course, the density matrix $\rho(\phi,\phi')$ with two surfaces carrying its field arguments. These surfaces {\em semiclassically} are the boundaries of either Euclidean or Lorentzian spacetime, depending on the relevant size of the scale factor. The entire saddle point solution for $\rho(\phi,\phi')$ consists respectively of the Euclidean spacetime interpolating between them or of the Euclidean spacetime between $\Sigma$ and $\Sigma'$, sandwiched between the two layers of the Lorentzian spacetime. These two layers interpolate from $\Sigma$ to the unprimed argument of the density matrix and from $\Sigma'$ to its primed argument and correspond in the density matrix to the chronological and anti-chronological evolution factors of the well-known Schwinger-Keldysh technique \cite{Schwinger-Keldysh} for expectation values in thermal field theory. When calculating the trace in the statistical sum in view of unitarity these two factors cancel out, and the only contribution to the statistical sum remains from the Euclidean domain between the Euclidean-Lorentzian transition surfaces $\Sigma$ and $\Sigma'$. These surfaces are uniquely determined from the condition of smooth periodicity in the Euclidean time on the compact $S^1$, or as two turning points of the Euclidean trajectory for $a(\tau)$.
\begin{figure}[h]
\centerline{\includegraphics[width=6cm]{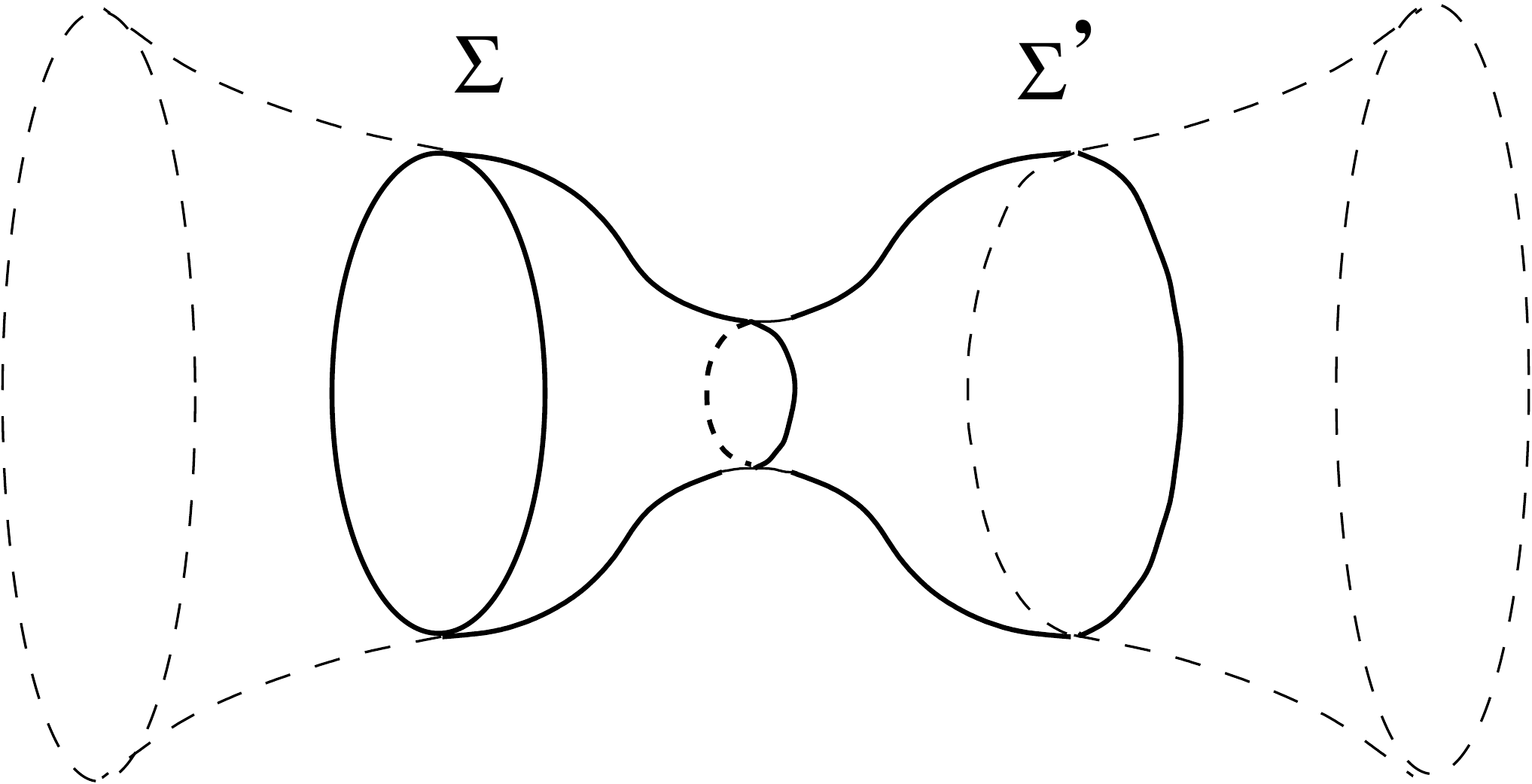}} \caption{\small
Density matrix instanton. Dashed lines depict the Lorentzian
Universe nucleating at minimal surfaces $\Sigma$ and $\Sigma'$.
 \label{Fig2}}
\end{figure}

The equations for these cosmological instantons have the form of the effective Friedmann equation in the Euclidean time $\tau$ ($\dot a=da/d\tau$),
    \begin{eqnarray}
    &&-\frac{\dot a^2}{a^2}+\frac{1}{a^2}
    -B \left(\,\frac{\dot a^4}{2a^4}
    -\frac{\dot a^2}{a^4}\right) =
    \frac\rho{3M_P^2}+\frac{C}{ a^4},\,\,                \label{efeq}\\
    &&C =
    \frac{B}2+\frac{R(\eta)}{3M_P^2},           \label{C}\\
    &&B=\frac{\beta}{8\pi^2M_P^2},               \label{B}
    \end{eqnarray}
where $M_P=1/\sqrt{8\pi G}$ is the reduced Planck mass, $\rho$ is the overall energy density of matter fields other than the conformal particles, $\beta$ is the coefficient of the Gauss-Bonnet term in the total conformal anomaly of these particles and $R(\eta)$ is their radiation energy density.\footnote{It should be emphasized that non-conformal matter was not completely excluded in the original setup and the gravitational sector of the theory was not assumed to be Wey invariant at all. In particular, the role of $\rho$ could be played by a fundamental cosmological constant, its particular value being selected from the existence of the periodic Euclidean saddle-point solution, as it was in the simplest model of \cite{Barvinsky:2006tu}. In more realistic models the role of $\rho$ is played by the non-conformal inflaton field in the slow roll regime or the scalaron field of the Starobinsky $R^2$-model \cite{EPJC,Hill}, see below. Moreover, $\rho$ can contain ordinary particle matter of negligible amount in the early Universe, but quantum created during inflation in view of its non-conformal nature and, therefore, starting to dominate at later stages of the evolution, see footnote \ref{creation}.} The latter is given by a boson or fermion sum over field modes with energies $\omega$ on a unit 3-sphere at the comoving temperature $1/\eta$ -- the inverse of the instanton circumference $\mathbb{S}^1$ measured in units of the conformal time,
    \begin{eqnarray}
    &&R(\eta)=\frac1{2\pi^2}\sum_{\omega}\frac\omega{
    e^{\omega\eta}\mp 1},\quad
    \eta=\int_{\mathbb{S}^1} \frac{d\tau}a.     \label{period}
    \end{eqnarray}

Note that $\gamma$ does not contribute to the above equations in view of conformal flatness of the FRW metric, while the coefficient $\alpha$ can always be renormalized to zero by a local $R^2$ counterterm, changes in $\alpha$ thus being equivalent to the inclusion of the non-minimally coupled scalaron of the Starobinsky model, see discussion in \cite{EPJC,Hill}.

The integro- differential equations (\ref{efeq})-(\ref{C}) form a bootstrap -- the amount of radiation constant $C$ is determined from (\ref{C}) by the underlying scale factor history $a(\tau)$ which, in its turn, is generated by the back reaction of this radiation on $a(\tau)$ via the effective Friedmann equation (\ref{efeq}). Their solutions represent the set of periodic $\mathbb{S}^3\times \mathbb{S}^1$ gravitational instantons\footnote{We use the term ``gravitational instanton" to avoid confusion with conventional instanton type solutions  which describe the interpolation between topologically  distinct but physically identical winding sectors $|k\ra$ in gauge theories. The corresponding periodic instantons (the so-called calorons with nontrivial holonomy) is the subject of the Appendix \ref{review} where we overview relevant for the present work results.} with the oscillating scale factor, {\em garlands} \cite{Barvinsky:2006uh,why}, that can be regarded as the thermal version of the Hartle-Hawking instantons \cite{noboundary}. When the matter density is constant or nearly constant and forms a "Hubble factor"
    \begin{eqnarray}
    H^2\equiv\frac\rho{3M_P^2}     \label{H^2}
    \end{eqnarray}
the scale factor oscillates $m$ times ($m=1,2,3,...$) between the maximum and minimum values, $a_-\leq a(\tau)\leq a_+$, so that the full period of the conformal time (\ref{period}) is the $2m$-multiple of the integral between the two neighboring turning points of $a(\tau)$, $\dot a(\tau_\pm)=0$. Similarly, the full period of the {\em proper} Euclidean time on these periodic $m$-fold {\em garland} instantons is given by the analogous integral,
    \begin{eqnarray}
    &&{\cal{T}}=\oint_{\mathbb{S}^1}d\tau=2m\int_{a_-}^{a_+}
    \frac{da}{\dot a}.                       \label{period1}
    \end{eqnarray}

These garland-type instantons exist only in the limited range of $H^2$ \cite{Barvinsky:2006uh}. As shown in \cite{Barvinsky:2006uh}, periodic solutions should necessarily belong to the domain
    \begin{eqnarray}
    B-B^2H^2\leq C\leq \frac1{4H^2}, \label{triangle}
        \end{eqnarray}
where they form a countable, $m=1,2,...$, sequence of one-parameter families interpolating between the lower and upper boundaries of this domain in the two-dimensional plane of $H^2$ and $C$. This sequence with $m\to\infty$ accumulates at the upper bound of $H^2_{\rm max}=1/2B$ (and minimal value of $C_{\rm min}=B/2$), which correspond to the bound on the effective cosmological constant
    \begin{eqnarray}
    \varLambda_{\rm max}=\frac{12\pi^2 M_P^2}\beta. \label{Lambdamax}
    \end{eqnarray}
The lower bound $H^2_{\rm min}$ -- the lowest point of $m=1$ family -- can be obtained numerically for any field content of the model.

For solutions close to the upper boundary of the domain (\ref{triangle}), $C\simeq1/4H^2$, the scale factor oscillates with a very small amplitude, and one can write down the following approximation
    \begin{eqnarray}
    &&a^2=\frac{1+\sqrt{1-4CH^2}\cos(\varOmega\tau)}{2H^2},\label{approxE}\\
    &&\varOmega=\!\frac{2H}{\sqrt{1-2BH^2}},
    \end{eqnarray}
which is valid for $\sqrt{1-4CH^2}\ll 1-2BH^2$ \cite{EPJC,Hill}. The full period of the $m$-folded instanton is thus
    \begin{eqnarray}
    {\cal{T}}=\frac{2\pi m}{\varOmega}=\frac{\pi m}H\,\sqrt{1-2BH^2}.     \label{T}
    \end{eqnarray}

Remarkably, the bootstrap equations (\ref{efeq})-(\ref{C}) have explicit solution for large $m$ and close to the upper boundary of the domain (\ref{triangle}) \cite{Barvinsky:2006uh}. In this limit the Hubble parameter is close to the upper bound of its range
    \begin{eqnarray}
    H^2\simeq\frac1{2B}
    \left(1-\frac{\ln^2 m^2}{2\pi^2m^2}\right),\quad m\gg 1, \label{Hm}
    \end{eqnarray}
corresponding to the maximal value of the effective cosmological constant (\ref{Lambdamax}).

We would like to make few comments on the physical meaning of the topological parameter $m$ which enters  eq. (\ref{period1}). This parameter looks very similar to the integer instanton number in gauge theories where the Euclidean path integral is defined as the sum over all topological sectors, so that it is tempting to consider summation over $m$. However,
$m$ is not an independent parameter of the cosmological instantons of the above type. Each instanton is parametrized by two dimensional parameters, $H^2=\varLambda/3$ -- the cosmological constant or the energy scale of the model and $M_P$ -- the Planckian mass or gravitational coupling constant. The folding number $m$ is in one to one correspondence with the energy scale $H$ as in (\ref{Hm}). Therefore, under a general assumption that at later times the cosmological models with different values of $H$ decohere and become observable, one should not sum over different values of $m$ in the contribution to the initial conditions for inflation with a given $H$. The concrete values of $H$ and $m$ should, thus, be selected by matching with observations.

Inflation stage in this model starts after the ``nucleation" of the system from the gravitational instanton when the evolution in the Lorentzian time begins. The Lorentzian time history of the scale factor $a_L(t)$ originates by the analytic continuation of the approximate solution (\ref{approxE}) to $\tau=2m\pi/\varOmega+it$. This leads to the replacement of oscillatory behavior of $\cos(\varOmega\tau)$ by exponentially growing $\cosh(\varOmega t)$, so that at later times nonlinear effects start dominating.  When solved with respect to $\dot a^2$ Eq.(\ref{efeq}) takes in the Lorentzian spacetime, $\dot a^2(\tau)=-\dot a_L^2(t)$,  the manifestly general relativistic form (cf. (\ref{efeqE})-(\ref{Hpm}) below),
    \begin{eqnarray}
    &&\frac{\dot a_L^2}{a_L^2}+\frac1{a_L^2}=
    \frac{\varepsilon}{3M^2_{\rm eff}(\varepsilon)},\quad
    \varepsilon=\rho+\frac{R}{a^4_L},               \label{AAA}\\
    &&M^2_{\rm eff}(\varepsilon)=
    \frac{M_P^2}2\left(\,1+\sqrt{1
    -\frac{\beta\,\varepsilon}{12\pi^2M_P^4}}\,\right),   \label{BBB}
    \end{eqnarray}
with the effective Planck mass $M_{\rm eff}(\varepsilon)$ depending on the full matter density $\varepsilon$ which together with $\rho$ includes the primordial radiation of the conformal cosmology.

As shown in \cite{EPJC,Hill}, the above Euclidean-Lorentzian scenario remains valid also when the matter density $\rho$ is represented by an appropriate potential of the slowly varying scalar field playing the role of the inflaton.  The evolution consists in the fast quasi-exponential expansion during which the primordial radiation gets diluted, the inflaton field and its density $\rho$ slowly decay by a conventional exit scenario and go over into the quanta of conformally non-invariant fields produced from the vacuum.\footnote{\label{creation}  A realistic model should contain a sector of non-conformal fields which can be negligible on top of conformal fields in the early Universe but eventually starts dominating in the course of cosmological expansion.} They get thermalized and reheated to give a new post-inflationary radiation with a sub-Planckian energy density, $\varepsilon\to\varepsilon_{\rm rad}\ll M_P^4/\beta$. Therefore, $M_{\rm eff}$ tends to $M_P$, and one obtains a standard general relativistic inflationary scenario for which initial conditions were prepared by the garland instanton of the above type.

Interestingly, this model can serve as a source of quantum initial conditions for the Starobinsky $R^2$-inflation \cite{R^2} and Higgs inflation theory \cite{Higgsflation,Higgsflation1}, in which the effective $H^2$ is generated respectively by the scalaron and Higgs field. In particular, the observable value of the CMB spectral tilt $n_s\simeq 0.965$ in these models can be related to the exponentially high instanton folding number \cite{EPJC,Hill}
    \begin{eqnarray}
    m\simeq\exp\frac{2\pi}{\sqrt{3(1-n_s)}}\sim 10^8,     \label{m}
    \end{eqnarray}
whereas the needed inflation scale in these models $H\sim 10^{-6}M_P$ determines the overall parameter $\beta\sim 10^{13}$ generated by a hidden sector of conformal fields \cite{Hill,CHS}. If this sector is built of higher spin conformal fields \cite{CHS}\footnote{Such a high value of $\beta$ cannot be reached with low spin conformal fields having $\beta=(1/180)\big(\mathbb{N}_0+11 \mathbb{N}_{1/2}+62 \mathbb{N}_{1}\big)$, unless the numbers $\mathbb{N}_s$ of fields of spin $s$ are tremendously high. On the contrary, this bound on $\beta$ can be reached with a relatively low tower of higher spin conformal fields, because partial contribution of spin $s$ to $\beta$ grows as $s^6$ and, moreover, this scaling guarantees that the theory with multiple quantum species remains deeply below its gravitational cutoff \cite{cutoff,cutoff1}.}, then the gravitational cutoff \cite{cutoff,cutoff1} of the model turns out to be several orders of magnitude higher than the inflation scale, which justifies the omission of the graviton loop contribution and the use of the above nonperturbative (trace anomaly based) method. This concludes our overview of the previous results  \cite{Barvinsky:2006uh,Barvinsky:2006tu,why,EPJC,Hill} which play an important role in constructions presented in the following sections.

\section{The topology as the source of the vacuum energy}\label{topology}
The goal here is to overview the basic ideas advocated in  \cite{Zhitnitsky:2013pna,Zhitnitsky:2014aja,Zhitnitsky:2015dia}.
  We explain a number of technical elements related to these ideas in Appendix \ref{review}, while
here we present the corresponding arguments using a simple plain language and analogies, see next subsection \ref{basics}.
The basic prescription of the vacuum energy which enters the Friedmann equations will be explained in subsection \ref{holonomy}.
In  subsection \ref{interpretation}  we list a number of key technical elements  of the proposal relevant for cosmological applications.
  \subsection{Intuitive picture}\label{basics}

The new paradigm advocated in \cite{Zhitnitsky:2013pna} is based on a fundamentally novel view on the nature and origin of the inflaton field which is drastically different from the conventional viewpoint that the inflaton is a dynamical local field $\Phi$.
In this new framework the inflation is a {\it genuine quantum effect} in which the role of the inflaton is played by an auxiliary topological field.
A similar field, for example, is known to emerge in the description of a topologically ordered condensed matter (CM) system realized in nature.
This field does not propagate, does not have a canonical kinetic term, as the sole role of the auxiliary field is to effectively describe the dynamics of the topological sectors of a gauge theory which are present in the system.
The corresponding physics is fundamentally indescribable in terms of any local propagating fields (such as $\Phi(x)$).
It might be instructive to get some intuitive picture for  the   vacuum energy  in this framework formulated in terms of a CM analogy.
 Such an intuitive picture is quite helpful in getting a rough idea about the nature of the inflaton in the  framework advocated in this work.

Imagine that we study the Aharonov-Casher effect.
We insert an external charge into a superconductor when  the electric field $E$ is screened, i.e. $E\sim Q\exp(-r/\lambda)$ with $\lambda $ being the penetration depth.
Nevertheless, a neutral magnetic fluxon will be still sensitive to an inserted external charge $Q$ at arbitrary large distances in spite of the screening of the physical field.
This genuine quantum effect is purely topological and non-local in nature and can be explained in terms of the dynamics of the gauge sectors which are responsible for the long range dynamics.
Imagine now that we study the same effect but in a  time dependent background.
The corresponding topological sectors which saturate the vacuum energy will be modified due to the external background.
However, this modification can not be described in terms of any local dynamical fields, as there are no any propagating long range fields in the system since the physical electric field is screened.
For this simplified example, the dynamics of the inflaton corresponds to the effective description of the modification of topological sectors when the external background slowly changes.
The effect is obviously non-local in nature as the Aharonov-Casher effect itself is a non-local phenomenon, and cannot be expressed in terms of $F_{\mu\nu}$.

One should emphasize that many crucial  elements of this proposal  have in fact been tested using the numerical lattice Monte Carlo simulations in strongly coupled QCD.
Furthermore,  this  fundamentally new sort of energy can be in principle studied in tabletop experiments by measuring some specific corrections to the Casimir pressure in the Maxwell theory, see  remarks and references in concluding section \ref{test}.
In next subsection we specifically list some important  technical elements which will be used in the construction.

\subsection{QCD holonomy mechanism of vacuum energy}\label{holonomy}
Let us now get to the discussion of the nature of the effective cosmological constant or a Hubble factor (\ref{H^2}) in the modified Euclidean Friedmann equation (\ref{efeq}). Our interpretation in the present work is based on the prescription that the relevant energy is in fact the difference $\Delta\rho\equiv\rho -\rho_{\mathrm{flat}}$ between the energies  of a system in a non-trivial background  and flat space-time geometry,  similar to the well known Casimir effect when the observed   energy is  a difference between the energy computed for a system with conducting boundaries   and infinite Minkowski flat space.  In this framework it is quite natural to define the ``renormalized vacuum energy'' to be zero in flat space-time  vacuum wherein the Einstein equations are automatically satisfied as the Ricci tensor identically vanishes.

In the present context  such a definition $\Delta\rho\equiv \rho_{\rm FLRW} -\rho_{\mathrm{Mink}}$ for the vacuum energy for the first time was advocated   in 1967   by Zeldovich~\cite{Zeldovich:1967gd} who argued that  $\rho_{\text{vac}}=\Delta\rho \sim Gm_p^6 $ must be proportional to the gravitational constant with $m_p$ being the proton's mass. Later on such  definition for the relevant energy $\Delta\rho\equiv \rho_{\rm FLRW} -\rho_{\mathrm{flat}}$ which  enters the Einstein equations has been advocated from   different perspectives in a number of papers written by the researches from different fields, including particle physics, cosmology, condensed matter physics,  see e.g.  relatively recent works~\cite{Bjorken:2001pe, Schutzhold:2002pr, Klinkhamer:2007pe,   Thomas:2009uh,Maggiore:2010wr}, and  review article \cite{Sola:2013gha} with
large number of   the original references.

This subtraction prescription is consistent with conventional  subtraction procedure of the divergent ultra local bare cosmological constant because in the infinitely large  flat space-time the corresponding contribution is proportional to the  $\delta^4(x)$ function, see (\ref{K1}). At the same time  the  nontrivial corrections to    $\Delta\rho$ are  non-local functions of the geometry and cannot be renormalized by any UV counter-terms.

Precisely this feature of non-locality implies that the relevant energy $\Delta \rho$ which enters the Friedmann equation, see  (\ref{Delta1}) below,  cannot be expressed in terms of a gradient expansion in any effective field theory. Additional arguemnts
supporting the same claim  on impossibility to formulate the relevant physics in terms of any local effective field, such as inflaton  $\Phi(x)$  will be presented in the following subsection \ref{interpretation}.

This prescription is also consistent with the renormalization group approach   advocated in  \cite{Shapiro:1999zt,Shapiro:2000dz,Sola:2013gha}. In  fact, it is direct consequence of the  renormalization group approach when we fix a physical parameter at one point of normalization to predict
its value at a different normalization point. In the present work with the geometry  $\mathbb{S}^3\times \mathbb{S}^1$, the proper length of the $\mathbb{S}^1$-period being $\cal{T}$,  it implies that  the vacuum energy in the  Friedmann  equation (\ref{efeq}) is  $\rho\equiv\rho({\cal{T}}^{-1})-\rho(0)$, where $\rho({\cal{T}}^{-1})$ is the energy of the gauge field holonomy on a compactified spacetime coordinate of length $\cal{T}$. It can be interpreted as the RG normalization point $\mu\sim  {\cal{T}}^{-1}$, where $T$ is the  size of the compactified Euclidean time dimension given by (\ref{period1}).  As we already mentioned, this prescription is consistent with the  Einstein equations when the vacuum energy  approaches zero,  $\Delta\rho\rightarrow 0$  for the flat  space-time   which itself may be considered as a limiting case with ${\cal{T}}\rightarrow\infty$.

Finally, with the expression for the energy of the gauge field holonomy winding across the compactified coordinate of the length ${\cal{T}}$  whose derivation we give in next subsection \ref{interpretation}, one has
\be
\label{Delta1}
\rho\equiv \rho_{\rm vac}[ \mathbb{S}^3
\times \mathbb{S}^1]-\rho_{\rm vac}[ \mathbb{R}^4  ]
 = \frac{ \bar{c}_{{\cal{T}}}\Lbar^3}{{\cal{T}}},
\ee
where $\Lbar$ is the scale of the underlying QCD-like gauge field theory and $\bar{c}_{{\cal{T}}}$ is some dimensionless ${\cal{O}}(1)$ constant whose precise value is not important for our argumentation.

Our final comment  in this subsection goes as follows. As we already mentioned the energy $\Delta\rho$ can be interpreted as a running cosmological constant within the renormalization group approach   advocated in  \cite{Shapiro:1999zt,Shapiro:2000dz, Sola:2013gha} with the only difference that odd powers of $H$ are also included into the series as a result of the IR sensitivity and non-locality   (in contrast with conventional UV renormalization) as discussed  in   Appendix \ref{review}.
The linear correction (which is a particular example of the odd power of $H$)  to the vacuum energy can be interpreted in terminology    \cite{Shapiro:1999zt,Shapiro:2000dz, Sola:2013gha} as possibility of running cosmological constant at very low $\mu\sim {\cal{T}}^{-1}\ll M_P$. This running is  originated from non-perturbative and non-local physics  in QFT (through the nontrivial holonomy along $ {\mathbb{S}^1}$) and can not be seen at any finite level in perturbation theory, as entire ``non-dispersive" vacuum energy can  not  be generated   in perturbation theory, see some technical comments on this matter in Appendix \ref{sec:holonomy}.

As we will   see in next subsection, the leading  correction to the vacuum energy (\ref{Delta1}) is in fact proportional to $H$, and this linear in $H$ correction in the effective Friedmann equation is saturated by the IR-sensitive topological configurations with nontrivial holonomy which cannot be expressed in terms of any local propagating degrees of freedom.

\subsection{\label{interpretation} ``Non-dispersive" vacuum energy.  Cosmological context.}
We define the ``non-dispersive" vacuum energy $E_{\rm vac}$ in gauge theory in conventional way in terms of the path integral, see Appendix \ref{contact}.
Precisely this  vacuum energy  enters all the relevant correlation functions, including  the topological susceptibility as defined by
(\ref{chi}).

1. From the arguments of Appendix \ref{contact} one can infer that the $\theta$- dependent portion of the vacuum energy $E_{\rm vac}(\theta)$ can not be identified with any   propagating degrees of freedom. Furthermore,  all effects are obviously non-analytical  in coupling constant $\sim \exp(-1/g^2)$ and can not be seen in perturbation theory. These arguments obviously suggest that there is no any local effective field $\Phi(x)$ (inflaton) which could describe
  these features of the vacuum energy in gauge theories. These arguments are obviously consistent with our  discussions in previous  subsection \ref{holonomy}.

2. One can view the relevant topological Euclidean configurations which satisfy the properties from item 1 above  as the 3d     magnetic monopoles  wrapping around  $\mathbb{S}^1$ direction. These configurations are characterized by the  non-vanishing holonomy (\ref{polyakov}), which eventually generate the linear correction $\sim1/{\cal{T}}$ to the vacuum energy density represented by eqs.(\ref{ratio1}) and (\ref{vacuum_energy1}) below.

3. In the cosmological context such configurations are highly unusual objects: they obviously describe the non-local physics
  as the holonomy (\ref{polyakov}) is a nonlocal object. Indeed,  the holonomy  defines the dynamics  along the entire history  of evolution of the system in the given confined phase: from the very beginning to the very end. There is no contradictions with causality    in the system as there is no any   physical degrees of freedom   to propagate along this   path,   see item 1 above. Furthermore,  this entire gauge configuration is a mere  saddle point in Euclidean   path integral computation which describes the instantaneous tunnelling event, rather than propagation of  a physical  degree of freedom capable to carry an information/signal.

 4. The generation of the ``non-dispersive" energy $E_{\rm vac}$ is highly non-local   effect. In particular, formulae  eqs.(\ref{ratio1}) and (\ref{vacuum_energy1}) below explicitly  show that small variations of the background produces
  large  linear correction $\sim {\cal{T}}^{-1}$ at small ${\cal{T}}^{-1}\rightarrow 0$       as a result of this non-locality.
  Precisely this feature of non-locality implies that the relevant energy $\Delta \rho$ which enters the Friedmann equation (\ref{Delta1}),   cannot be expressed in terms of a gradient expansion in any effective local field theory as emphasized in section \ref{holonomy}.

 5. Our subtraction prescription as explained in section \ref{holonomy} is consistent will all fundamental principles  of QFT.
 What is more important is that the correction   to the energy $\Delta \rho$ which enters the Friedmann equation (\ref{Delta1}),  cannot be renormalized by any UV counter-terms as it is generated by non-local configurations.

  6. The basic  assumption of this work is that the same pattern (as highlighted in items 1-5 above) holds for other manifolds.
   In other words, we assume that the vacuum energy density for $\mathbb{S}^3 \times \mathbb{S}^1$ manifold receives a linear correction ${\cal{T}}^{-1}$ in comparison with flat   $\mathbb{R}^3 \times \mathbb{S}^1$ geometry, similar to the computations in hyperbolic space $\mathbb{S}^1\times\mathbb{H}^3$ where computations can be explicitly  performed, as reviewed in Appendix \ref{sec:holonomy},  i.e.
      \be
\label{ratio1}
\frac{E_{\rm vac}[ \mathbb{S}^3 \times \mathbb{S}^1]}{E_{\rm vac}[ \mathbb{R}^3 \times \mathbb{S}^1]}
 \simeq  \left(1-   \frac{c_{{\cal{T}}}}{{\cal{T}}\Lbar} \right),
\ee
where $c_{{\cal{T}}}$ is a coefficient of order one, similar to computations in Appendix \ref{sec:holonomy}. Formula (\ref{ratio1}) plays the crucial role in our  arguments in sections  \ref{Inflation} and \ref{Inflation-2}.

One can use conventional thermodynamical relation
\be
\label{thermodynamics}
dF=TdS-PdV, ~~~~ P=-\frac{\partial F}{\partial V}|_S
\ee
to convince yourself that the correction $\sim {\cal{T}}^{-1}$ does not modify the  equation of state. In fact, it behaves exactly in the same way  as the cosmological constant does,  i.e.
\be
\label{EoS}
P&=&-\frac{\partial F}{\partial V}=+ \frac{32\pi^2}{g^4} \Lbar^4  \left(1-   \frac{c_{{\cal{T}}}}{{\cal{T}}\Lbar} \right)\nonumber\\
\rho&=&\frac{F}{V}= - \frac{32\pi^2}{g^4} \Lbar^4  \left(1-   \frac{c_{{\cal{T}}}}{{\cal{T}}\Lbar} \right),
\ee
where we use formula (\ref{Z3}) for $F$ with correction factor (\ref{ratio1}). The correction $\sim {\cal{T}}^{-1}$ does not modify the equation of states $w=-1$, which   is normally associated with  the  cosmological constant contribution,
\be
\label{omega}
w\equiv\frac{P}{\rho}=-1.
\ee
Finally, using (\ref{ratio1}) the vacuum energy for $ \mathbb{S}^3 \times \mathbb{S}^1$ manifold can be represented as follows
\be
\label{vacuum_energy1}
&&E_{\rm vac}[ \mathbb{S}^3 \times \mathbb{S}^1]\simeq   -\frac{32\pi^2}{g^4} \Lbar^4  \left(1-   \frac{c_{{\cal{T}}}}{{\cal{T}}\Lbar} \right)\nonumber\\
&&\simeq -\frac{32\pi^2}{g^4} \Lbar^4 +  \Lbar^3 \frac{\bar{c}_{{\cal{T}}}}{{\cal{T}}}+{\cal{O}}(\frac{1}{{\cal{T}}^2}), .
\ee
where we redefined $\bar{c}_{{\cal{T}}}\equiv\frac{32\pi^2}{g^4}c_{{\cal{T}}}$ as the parameter $c_{{\cal{T}}}\sim 1 $ is expected to be order of one (based on the previous experience) but is not yet  known.

We conclude this section with few important  comments which are relevant for  the physical interpretation of the obtained  results.

$\bullet$ All computations presented above, as usual,  are performed in the Euclidean space-time where the relevant gauge configurations describing the tunnelling processes are defined. Using this technique we computed the energy  density $\rho$ and the pressure $P$ in the Euclidean space.
As usual, we assume that there is analytical continuation to Lorentizan space-time where the physical energy density has  the same form.
This is of course, conventional procedure for the QCD practitioners who normally perform computations on the lattice using the Euclidean formulation, while the obtained results are expressed  in physical terms in Minkowski  space-time.  In our context it means that the parameters $P, \rho$ and equation of state (EoS) as given by (\ref{omega}) are interpreted as the corresponding parameters in physical Lorentizan  space-time.

$\bullet$ Therefore, the driving force for the deSitter behaviour  in the Lorentzian space is not a local dynamical inflaton field $\Phi(x)$ which never emerges in our framework.  Rather the driving
force in our scenario should be thought as a Casimir type vacuum energy which is generated by numerous tunnelling transitions in a strongly coupled gauge theory determined by the  dimensional parameter $\Lbar$. Precisely this parameter replaces the dimensional parameters
from inflaton potential $V[\Phi(x)]$ which  cosmology practitioners normally use in their studies.

$\bullet$ The equation of state (\ref{omega}) in Lorentizan  space-time obviously implies the deSitter expansion.  The corrections due to the radiation $\rho_r$ and matter $\rho_m$ can be easily incorporated into the Friedmann equation written in Lorentizan  space-time. The interaction of the system with standard model (SM) particles will modify the EoS (\ref{omega}). Precisely these modifications  to EoS (\ref{omega})  will be responsible for the end of inflation as described in section \ref{inflation_end}.

\exclude{
$\bullet$ All computations, including the analysis of the cosmological perturbations
can be entirely formulated in terms of parameters $P$ and  $\rho$ as advocated  in \cite{mukhanov}. Precisely this description in terms of  $P$ and  $\rho$   avoids  any mentioning of any local physical propagating fields (such as inflaton $\Phi (x)$) will be exploited in section  \ref{inflation_end}.
}

\section{The holonomy Inflation. Model-1.}\label{Inflation}
The origin of inflation in the model reviewed  in previous section \ref{sec:grav-instanton}  is based on two important ingredients -- the vacuum energy (\ref{H^2}) of a certain local nature and the hidden sector of conformal fields critically important for the contribution of the conformal anomaly and generation of the thermal radiation in effective Friedmann equation. Key technical element for the successful inflation is the presence of   $\mathbb{S}^1$ which emerges  in the system as a result of thermal initial state formulated in terms of the density matrix. We keep this first ingredient of our construction from previous studies  as will be explained below in subsection \ref{S_1}.

A new idea which is advocated in the present work is that the second  important ingredient of this framework, the vacuum energy,  may  be originated from some nontrivial    non-local gauge configurations. This structure in our proposal is fundamentally different from all conventional inflationary models because this source of the vacuum energy
   cannot be  expressed in terms of any  local degrees of freedom such  as scalar inflaton $\Phi(x)$.
\exclude{
Remarkably, this model opens a prospect of another mechanism of the vacuum energy, which is based not on local field-theoretical degrees of freedom like $R^2$ or scalar inflaton,
but rather relies on nontrivial topological properties of the configuration space of gauge fields. }

In our construction this source of the vacuum energy is generated by the gauge configurations with
 nontrivial holonomy in the QCD-like field theory as explained in subsection  \ref{holonomy}.
 This construction uses  exactly   the nontrivial topology $\mathbb{S}^1\times \mathbb{S}^3$ of the gravitational instanton considered above. In its turn, the origin of this topology -- compactification of the Euclidean time on a circle $\mathbb{S}^1$ -- is entirely due to a subtle effect of conformal radiation, whereas the inflation compatible value of the vacuum energy is the effect of this holonomy in the QCD-like gauge theory with a subplanckian scale as explained in subsection \ref{inflation-1}.

We treat  {\it Model-1} considered in this section as a toy model where, one one hand,  one can demonstrate all the crucial elements of the construction.
 On    other hand,  one can adjust parameters in a such a way that all computations are under complete theoretical control and the  semiclassical approximation is justified. Unfortunately, this model is not very natural as it requires very large instanton folding number $m$ and very large $\beta$ to be consistent with observations.

 In next section \ref{Inflation-2} we consider {\it Model-2}, which  is naturally consistent with all presently available observations without special selection of the parameters $\beta$ or $m$. However, we should relax some technical requirements for Model-2 in which case  the semiclassical approach   is not formally justified.  
 \exclude{Nevertheless, all crucial elements of the construction are in place such the model deserves   further  serious analysis,  including  comparison with future precise measurements such as the spectral index $n_s$, the tensor fraction $r_T$, the tensor tilt $n_t$, the running of the spectral index
 $\partial n_s/\partial \ln k$,  the running of the tensor tilt,  $\partial n_t/\partial \ln k$.
}

\subsection{The effect of the radiation generating $\mathbb{S}^1$}\label{S_1}
The effect of the radiation related to the difference $C-B/2$ in (\ref{C}) is indeed quite subtle because the radiation itself is strongly suppressed. For a high folding number $m\gg 1$ according to equation (\ref{Hm}) it is  proportional to
    \begin{eqnarray}
    C-\frac{B}2\simeq B\left(\frac{\ln m}{\pi m}\right)^2
    \end{eqnarray}
and very small for instanton solutions at the tip of the triangular domain (\ref{triangle}) with $H^2\simeq 1/2B$ and $C\simeq 1/4H^2$. At the same time   merely the existence of the radiation enforces us to consider the topology $\mathbb{S}^3\times \mathbb{S}^1$. If one ignores the radiation then the topology $\mathbb{S}^3\times \mathbb{S}^1$ reduces to $\mathbb{S}^4$.   This easily follows from the effective Friedmann equation (\ref{efeq}) with $\rho=3M_P^2H^2$ when it is cast, by solving it with respect to $\dot a^2$, into the form:
    \begin{eqnarray}
    &&-\frac{\dot a^2}{a^2}+\frac1{a^2}={\cal H}^2(a), \label{efeqE}\\
    &&{\cal H}^2(a)\equiv
    \frac1B\,
    \left(1-\sqrt{1-2BH^2-\frac{2BR}{3M_P^2a^4}}\right).  \label{Hpm}
    \end{eqnarray}
For $R=0$ it gives as a solution the Euclidean sphere,
$a(\tau)=\sin({\cal H}\tau)/{\cal H}$, of the radius $1/{\cal H}=(1/2H^2)(1-\sqrt{1-2BH^2})$, whereas any however small amount of radiation would provide a bouncing of $a$ back from some nonzero minimal value, otherwise $a=0$ occurring at the pole of $\mathbb{S}^4$.

But the contribution of such spherical (Hartle-Hawking) instantons to the path integral  is completely suppressed as argued in \cite{Barvinsky:2006uh,Barvinsky:2012qm}. Technically, this suppression occurs as a result of  the conformal anomaly which changes the sign of the negative classical action on $\mathbb{S}^4$ and, moreover, makes it divergent at the poles of the 4-sphere at $a\rightarrow 0$. Thus, it is entirely due to the radiation of conformal particles, that the scale factor never shrinks to zero, which allows one to compactify time on a circle and get the $\mathbb{S}^3\times \mathbb{S}^1$ topology, which can bear a nontrivial gauge field holonomy.

\subsection{QCD holonomy and inflation scale}\label{inflation-1}
The prescription we are advocating in the present work   essentially corresponds to the identification of the vacuum energy (\ref{Delta1}) with the energy density $\rho$ in the Hubble factor $H^2$ (\ref{H^2}) of the effective Friedmann equation (\ref{efeq}), i.e.
\be
\label{consistency}
H^2=\frac\lambda{{\cal{T}}},\quad
\lambda\equiv\frac{\bar{c}_{{\cal{T}}}\Lbar^3}{3M_P^2}.
\ee
With the instanton period of $m$-folded garland (\ref{T}), which is inverse proportional to $H$, this immediately gives
    \be
    H\sqrt{1-2BH^2}=\frac\lambda{\pi m}.       
    \label{consistency0}
    \ee
    This equation is always correct for any value of  $BH^2$. 
    
    However, the bootstrap self-consistency solution 
always has the property that  $2BH^2={\cal{O}}(1)$ as shown in detail in 
previous papers on CFT driven cosmology. The corresponding results will be discussed at the end of this subsection, while now we want to make few comments related to the the small value of parameter  $BH^2\ll 1$, which can be achieved in Model-2, to be discussed in the  next section \ref{Inflation-2}.

   If  we formally take $BH^2\ll 1$ in expression (\ref{consistency0})
the term  $BH^2\ll 1$ can be numerically neglected   in which case    $\lambda\sim H$ and the  equation (\ref{Delta1})
    assumes the form
    \be
\label{Delta2}
\rho(H)\equiv \rho_{\rm vac}[ \mathbb{S}^3
\times \mathbb{S}^1]-\rho_{\rm vac}[ \mathbb{R}^4  ]
 \simeq H\,\frac{  \bar{c}_{{\cal{T}}}\Lbar^3 }{\pi m},
\ee
    which explicitly shows the linear dependence of the vacuum energy on the Hubble constant, $\rho(H)\sim H$, as previously  claimed.
One can   see from (\ref{ratio1}),  (\ref{vacuum_energy1}) that  the source of this linear correction   to the vacuum energy
   is related to the term proportional to ${\cal{T}}^{-1}$ which represents the  inverse  size of $\mathbb{S}^1$ manifold for our geometry, and    proportional to  $H$
   in our framework. Needless to say that this linear (with respect to $H$) correction is  saturated by the IR topological configurations with nontrivial holonomy
   which cannot be expressed in terms of any local propagating fields as explained in Appendix \ref{review}.
   Therefore, this term  cannot be written in  a  conventional gradient expansion in an effective field theory as it represents a global, rather than local characteristic of the system.

Our next step is to make these computations   self-consistent   satisfying  the semi-classicality condition.
Formally, this condition is expressed as the bootstrap equation with solution (\ref{Hm}). The physical meaning of the  enforcement of  the
bootstrap equation as explained in previous section \ref{sec:grav-instanton} and original papers  \cite{Barvinsky:2006uh,Barvinsky:2006tu,why}
is that the temperature of the system (therefore the size of   $\mathbb{S}^1$) cannot be an arbitrary parameter.
 Instead, it  must be determined by the system itself.
In other words, the size of the manifold changes as a result of accounting for the feedback to adjust the changes of the vacuum energy.
This formal enforcement obviously implies that all dimensional parameters must be of  order of $M_P$ as the only scale of the problem.
 The deviation from the Planck  scale may only occur
if some very small or very large dimensionless parameters are present in the system.
In our Model-1 there are two  free parameters,  $\beta$, which effectively counts the number of degrees of freedom,  and the instanton number $m$ which, in principle,   assumes any value.

In this section,  in Model-1,  we want to proceed with self consistent computations. Therefore, we enforce the semi-classicality conditions. In this case,
for large $m$ and the value of $H$ determined  by the bootstrap solution (\ref{Hm}) this equation gives the expression for the parameter $\lambda$  which is equivalent to $\Lbar$, i.e.
    \be
    &&\lambda\simeq\frac{\ln m}{\sqrt B}
    =\sqrt{8\pi^2}\,\frac{M_P}{\sqrt{\beta}}\,\ln m,\\
    &&\Lbar\simeq \left(\frac{6\sqrt2\pi\ln m}{\bar{c}_{{\cal{T}}}} \right)^{1/3}
    \frac{M_P}{\beta^{1/6}}.                        \label{LambdaQCD}
    \ee
    As this model is considered to be a toy model, we can take $\beta$ as a free parameter and consider $\beta\gg 1$ such that
    \be
    \label{H}
    \frac{\Lbar}{M_P}\sim\frac{1}{{\beta}^{1/6}}\ll 1, ~~~~~~~
    \frac{H}{M_P}\simeq \frac{2\pi  }{{\beta}^{1/2}}\ll 1.
    \ee
    The key observation we want to make here  is that both parameters, $H$, and $\Lbar$ belong to the subplanckian scale according to (\ref{H}),
 which justifies the use of the semiclassical expansion discarding a negligible  contribution of graviton loops\footnote{Subplanckian scale of the model does not imply, however, that the $B$-terms in (\ref{efeq}) quadratic in curvature and generated by the conformal anomaly can be discarded. Effective action generating the conformal anomaly is nonlocal and strong in the infrared which is an artifact of the conformal invariance of the matter fields. In contrast to the quantum loops of conformal non-invariant graviton, loop effects of conformal matter are not suppressed by inverse powers of the Planck mass and their gravitational effect is treated beyond perturbation theory.}. Furthermore,
 there is a hierarchy of scales which parametrically holds for large $\beta\gg 1$:
 \be
 \label{hierarchy1}
 H\ll \Lbar\ll M_P.
 \ee
This hierarchy of scales once again demonstrates the self consistency of the computations (on the gauge side) because the ``non-dispersive" vacuum energy   (\ref{Delta1})  related to the holonomy is only generated in the  confined phase of the gauge  $\qcd$ theory at temperature  below $\Lbar$, which is automatically satisfied as a result of the hierarchy (\ref{hierarchy1}).

 Inflation scenario in the Lorentzian domain described in Sect.II (Eqs. (\ref{AAA})-(\ref{BBB}) above) holds also in the Gauge Holonomy Inflation model advocated in the present work. However the exit from inflation takes place via a decay of $H$ due to {\em helical} instability to be discussed below.  As mentioned above, if one attempts to match the parameters $m$ and $\beta$ with observational numbers, one should take extremely high values of these parameters. Indeed, the model becomes phenomenologically compatible with the CMB data within the Starobinsky $R^2$ or Higgs inflation theory when  the scale $H\sim 10^{-6}M_P$. It can be    generated by the SLIH scenario \cite{Barvinsky:2006uh} with $\beta\simeq 10^{13}$. In particular, it matches the observable value of the spectral tilt $n_s\simeq 0.97$ when the number of instanton folds equals (\ref{m}), $m\sim 10^8$.  If we assume that instead of the $R^2$-mechanism or the Higgs potential the vacuum energy is entirely due to the $\qcd$ holonomy mechanism of the above type, then from (\ref{LambdaQCD}) it follows that
\be
&&\Lbar\sim 0.05 M_P, ~~~ H\sim 10^{-6}M_P. \label{Lbarvalue}
\ee
The necessity to have very high value of $\beta$, which now can only be generated by a large hidden sector of conformal higher spin fields \cite{CHS}, makes this model rather speculative even though it justifies semiclassical expansion below the gravitational cutoff of \cite{cutoff,cutoff1}. Therefore we consider the second, much more natural   model without any hidden sectors filled by  large number of conformal fields.

\section{The holonomy Inflation. Model-2.}\label{Inflation-2}

The starting point in this section is the same set of equations discussed in previous sections. However, in (\ref{efeq}) we now ignore the higher derivative terms $\sim B \dot{a}^2$ and $\sim B\dot{a}^4$. It corresponds to disregarding  the higher derivative terms in the effective action as the typical scales of the problem will be much lower than the Planck scale $M_P$. The corresponding set of equations has been reviewed above, but now we consider the limit $BH^2\ll 1$ and for convenience of the readers repeat  some important formulae below.

\subsection{Overview of gravitational instanton solution}
The  scale  factor $a(\tau)$  oscillates between the maximum and minimum values $a_{\pm}$ determined as follows
 \be
 \label{a}
 a^2_{\pm}=\frac1{2H^2}\left(1\pm \sqrt{1-4CH^2} \right), ~~ a_{\pm}\equiv a(\tau_{\pm}), ~~ \frac{\varLambda}{3}\equiv H^2\nonumber
 \ee
The solution for the scale factor $a(\tau)$ is also known
 \be
 \label{a1}
 a^2(\tau)= \frac1{2H^2} \left(\,1- \sqrt{1-4CH^2}\,\cos \left(2H\tau \right) \,\right). ~~
 \ee
  \exclude{

  The full period of the conformal time $\eta$ is determined by the integral
  \be
  \label{eta}
  \eta=2k \int^{\tau_+}_{\tau_-}\frac{N(\tau)d\tau}{a(\tau)},~~ \tau_-=0, ~~ \tau_{+}= \frac{\pi}{2}\sqrt{\frac{3}{\Lambda}}~~~
  \ee
   where $k$ is integer number corresponding to generalization of the one gravitational instanton solution to the so-called ``garlands" representing its ``k" copies as discussed in \cite{Barvinsky:2006uh,Barvinsky:2006tu,Barvinsky:2012qm}.

     The novel idea which we are advocating in the present work is a different view (in comparison with \cite{Barvinsky:2006uh,Barvinsky:2006tu,Barvinsky:2012qm}) on the nature of the cosmological constant $\Lambda$  entering
    the modified Euclidean Friedmann equation (\ref{Friedmann}). Our interpretation in the present work  is based on
    the prescription that
  the relevant energy which enters the Friedmann  equations is in fact
 the difference $\Delta E\equiv E -E_{\mathrm{flat}}$ between the energies  of a system in a non-trivial background  and flat space-time geometry,  similar to the well known Casimir effect when the observed   energy is  a difference
  between the energy computed for a system with conducting boundaries   and infinite Minkowski flat space.  In this framework it is quite natural to define the ``renormalized vacuum energy'' to be zero in flat space-time  vacuum wherein the Einstein equations are automatically satisfied as the Ricci tensor identically vanishes.

  In the present context  such a definition $\Delta E\equiv (E_{\rm FLRW} -E_{\mathrm{Mink}})$ for the vacuum energy for the first time was advocated   in 1967   by Zeldovich~\cite{Zeldovich:1967gd} who argued that  $\rho_{\text{vac}}=\Delta E \sim GM_p^6 $ must be proportional to the gravitational constant with $M_p$ being the proton's mass. Later on such  definition for the relevant energy $\Delta E\equiv (E_{\rm FLRW} -E_{\mathrm{flat}})$ which  enters the Einstein equations has been advocated from   different perspectives in a number of papers written by the researches from different fields, including particle physics, cosmology, condensed matter physics,  see e.g.  relatively recent works~\cite{Bjorken:2001pe, Schutzhold:2002pr, Klinkhamer:2007pe,   Thomas:2009uh,Maggiore:2010wr}, and  review article \cite{Sola:2013gha} with
   large number of   the original references.

   This subtraction prescription is consistent with conventional  subtraction procedure of the divergent ultra local bare cosmological
   constant because in the infinitely large  flat space-time the corresponding contribution is proportional to the  $\delta^4(x)$ function, see (\ref{K1}). At the same time  the  nontrivial corrections to    $\Delta E$ are  non-local functions of the geometry and cannot be renormalized by any UV counter-terms.

  This prescription is also consistent with
the renormalization group approach   advocated in  \cite{Shapiro:1999zt,Shapiro:2000dz,Sola:2013gha}. In  fact, it is direct consequence of the  renormalization group approach when we   fix a physical parameter at one point of normalization to predict
its value at a different normalization point. In the present work with the geometry  $\mathbb{S}^3\times \mathbb{S}^1$  it implies that   the vacuum energy which enters the  Friedmann  equations is  $\Delta E\equiv [E(H)  -E_{\mathrm{flat}}]$. It is defined at    normalization point $\mu\sim H$, where the Hubble constant $H$ can be  expressed in terms of the cosmological constant $H=\sqrt{\Lambda/3}$ or in terms of the  size of the $\mathbb{S}^1$ manifold    which equals to
\be
\label{tau}
 \tau_{\rm total}=\oint_{\mathbb{S}^1} d\tau=   2(\tau_+-\tau_-)=\frac{2\pi}{H}=2\pi\sqrt{\frac{3}{\Lambda}}
\ee
 according to (\ref{eta}).  As we already mentioned, this prescription is consistent with the  Einstein equations
when the vacuum energy  approaches zero,  $\Delta E\rightarrow 0$  for the flat  space-time   which itself may be considered as a limiting case with $H\rightarrow 0$.

Our final comment  in this subsection goes as follows. As we already mentioned the energy $\Delta E$ can be interpreted as a running cosmological constant within the renormalization group approach   advocated in  \cite{Shapiro:1999zt,Shapiro:2000dz, Sola:2013gha} with the only difference that odd powers of $H$ are also included into the series as a result of the IR sensitivity and non-locality   (in contrast with conventional UV renormalization) as discussed  in   Appendix \ref{review}.
The linear correction (which is a particular example of the odd power of $H$)  to the vacuum energy
   can be interpreted in terminology    \cite{Shapiro:1999zt,Shapiro:2000dz, Sola:2013gha} as possibility of running cosmological constant at very low $\mu\sim H\ll m_p$. This running is  originated from non-perturbative and non-local physics  in QFT (through the nontrivial holonomy along $ {\mathbb{S}^1}$) and can not be seen at any finite level in perturbation theory, as entire ``non-dispersive" vacuum energy can  not  be generated   in perturbation theory, see some technical comments on this matter     in Appendix \ref{sec:holonomy}.

   In fact, one can   see from (\ref{ratio1}),  (\ref{vacuum_energy1}) that  the leading  correction to the vacuum energy
   is explicitly proportional to ${\cal{T}}^{-1}$ which represents the  inverse  size of $\mathbb{S}^1$ manifold for our geometry, and numerically    equals to  $H/(2\pi)$ according to (\ref{tau}). This linear with respect to $H$ correction is  saturated by the IR topological configurations with nontrivial holonomy
   which cannot be expressed in terms of any local propagating fields.
     }

Now we  implement the ideas formulated in the previous subsection. To proceed with this task we
identify the energy  (\ref{vacuum_energy1}) with the vacuum energy   entering  the Friedmann equations  as we discussed in previous section, i.e.
\be
\label{Delta}
\rho\equiv \rho_{\rm vac}[ \mathbb{S}^3 \times \mathbb{S}^1]- \rho_{\rm vac}[ \mathbb{R}^4  ]
 =  \frac{ \bar{c}_{{\cal{T}}}\Lbar^3}{{\cal{T}}}.
\ee
The prescription we are advocating in the present work   essentially corresponds to the identification of the vacuum energy (\ref{Delta}) with the cosmological constant $\Lambda/3$ entering the equation (\ref{H^2}), i.e.
\be
\label{consistency1}
3M_P^2\,H^2=\rho=\frac{\bar{c}_{{\cal{T}}}\Lbar^3}{\cal{T}}.
\ee
Up to this point the equation (\ref{consistency1}) identically coincides with our analysis in Eqs. (\ref{Delta1})-(\ref{consistency}) from the previous section.

 \subsection{Relaxing the semiclassical approximation}

New element for the Model-2 is as follows. We relax the bootstrap-like equation  and its  solution  (\ref{Hm}) for this model.
Essentially we unlink few parameters which were previously tightly linked. In particular, the de Sitter temperature being expressed in
terms of the size of $\mathbb{S}^1$ is unambiguously fixed by the radiation parametrized by parameter $\beta$.  This relation essentially
fixes the size  of $\mathbb{S}^1$ which is generated by the radiation and determined by the back reaction of $\mathbb{S}^1$ to the
corresponding radiation. The size of  $\mathbb{S}^3$    is also
not a free parameter in  semiclassical gravitational instanton solution. Essentially, by relaxing these links we assume that there could be another physics
which determines the size of the gravitational instanton (or a complicated network of strongly interacting gravitational instantons). A self-consistent  semiclassical approximation is obviously cannot be justified when some parameters enter from different physics. In Appendix \ref{inst-quark} we overview  a well-known example in strongly coupled gauge theory where the holonomy (and corresponding size of the manifold) is not fixed by hands, but rather is determined dynamically by strong quantum fluctuations. We suspect that a similar physics may emerge here.

In any case, for Model-2 we unlink the size of $\mathbb{S}^1$ from the radiation and treat it as a free parameter. To simplify our formulae, we also assume the lowest possible instanton number $m=1$ in all expressions in this section, which should be contrasted with our studies in previous section analyzing the Model-1 where  the  consistent description exists only for very large $m\sim 10^8$. This simplification does not modify our main results as the instanton number always accompanies  by dimensional parameter $\Lbar$
and dimensionless coefficient $\bar{c}_{{\cal{T}}}$ which are not yet known and can be always redefined\footnote{It does not imply that the system suffers some  ambiguities. In fact, the coefficient $\bar{c}_{{\cal{T}}}$ can be in principle computed from the first principles, while
the observation of the tensor fraction $r$ would unambiguously fix the relation between $H$ and $\Lbar$, see section \ref{inflation_end} with details. If both these parameters  were known, the  instanton number $m$ saturating the path integral can be also computed.}.

With these preliminary remarks, and  after substituting ${\cal{T}}=\pi/H$ (which is a good approximation in the regime  $CH^2\ll1$ we are interested in, see below)    the equation (\ref{consistency1}) 
can be rewritten in the following form
\be
\label{consistency2}
3M_P^2\, H^2=\frac{\bar{c}_{{\cal{T}}}\, \Lbar^3}{\pi}\,H.
\ee
This equation is very important  as it relates  the Hubble constant $H$ for our Euclidean geometry $\mathbb{S}^3 \times \mathbb{S}^1$ with
the vacuum energy generated by the gauge configurations with nontrivial holonomy,
\be
\label{Hubble}
H=\frac{\bar{c}_{{\cal{T}}}\Lbar^3}{3\pi M_P^2}, ~~~~ \rho=\frac{\bar{c}_{{\cal{T}}}^2\Lbar^6}{3\pi^2 M_P^2}
\ee
Few comments are in order. First of all,
the hierarchy of scales (\ref{H}), (\ref{hierarchy1}) characterizes the Model-1 from the previous section still holds in the present case
\be
 \label{hierarchy2}
 H\ll \Lbar\ll M_P.
 \ee
However, in Model-2 the hierarchy emerges not as  result of extremely large parameter $\beta\gg 1$, but rather,
as a result of new scale of the problem, $\Lbar$ which is a free dimensional  parameter of the system   generated by  the dimensional transmutation in classically conformal field theory and plays the same role in $\qcd$ as $\Lambda_{\rm QCD}\simeq 170$~MeV  plays  in QCD physics.
 \be
    \label{hierarchy3}
    \frac{H}{M_P}\sim \frac{\bar{c}_{{\cal{T}}}}{3\pi}\left(\frac{\Lbar}{M_P}\right)^3\ll 1, ~~~~~~~
    \frac{\Lbar}{M_P} \ll 1.
    \ee
 Parameter $ \Lbar/M_P\ll 1 $ plays the same role in Model-2 as parameter $\beta^{-1/6}\ll  1$ plays  in Model-1 as expressed by eq. (\ref{H}).
 The crucial difference, however, is that we unlink the size of $\mathbb{S}^1$ from the radiation by treating $\Lbar$ as a free dimensional parameter
 which defines a new gauge theory  coined as $\qcd$. It is assumed\footnote{We would like to make a short comment here why and how such unlink between these two parameters may occur.
 In weakly coupled semiclassical approximation in Model-1 the two parameters (the intensity of radiation characterized by the size of $\mathbb{S}^1$, 
which in its turn depends on the anomaly parameter $\beta$ in view of the 
bootstrap equation) are tightly  linked. In strongly coupled gauge theory as reviewed in Appendix \ref{inst-quark}  the holonomy and size of effective
$\mathbb{S}^1$ is determined dynamically. This is precisely the reason why   these two parameters in strongly coupled regime are not linked.
As reviewed in Appendix \ref{inst-quark} it is believed that in strongly coupled QCD the holonomy is also determined by the dynamics, the so-called ``confining holonomy" when the instanton dissociates into  $N$ constituents. Such a  phenomenon   may only occur  for topological configurations with nontrivial holonomy (\ref{holonomy1}). The known dependence of the vacuum energy on $\theta$ as $\cos(\frac{\theta}{N})$ is an explicit manifestation of the same nontrivial holonomy. }
 at this point that the size $\mathbb{S}^1$ where the holonomy is defined
 is determined by a different physics as discussed in Appendix \ref{inst-quark}.

\subsection{Subtle effects of the  radiation}
Due to the  hierarchy of scales mentioned in the previous subsection, one can explicitly check that the relevant parameter $\epsilon\equiv 4CH^2$ entering Eq.(\ref{a1}) is very small,
\be
\label{hierarchy}
\epsilon\equiv 4CH^2\sim \left(\frac{\Lbar}{M_P}\right)^6 \ll 1.
\ee
Indeed, as it follows from Eq.(\ref{a1}),  $\epsilon\leq 1$ because for larger $\epsilon$ the turning points disappear and monotonically changing $a(\tau)$ cannot form a periodic solution -- the saddle point of the partition function path integral. Thus in view of (\ref{C}) the amount of radiation $R(\eta)$ is always bounded from above -- though the Universe is born not in the vacuum state it is still essentially cold. The hottest possible Universe corresponding to a maximal value $\epsilon=1$ and minimal $\eta=\pi\sqrt{2}$ has a moderate maximal value of $R(\eta)=O(1)$. Actual smallness of $\epsilon$ assumed above follows from a subplanckian value of $H\ll M_P$, because Eq.(\ref{C}) is then equivalent to $\epsilon=\big(\beta+O(1)\big)H^2/4\pi^2M_P^2\ll 1$.

Thus one can simplify the formula  (\ref{a1}) and present approximate solution for $a(\tau)$ in the following form
 \be
 \label{a+}
  a(\tau)\simeq  \frac1H\,|\sin (H\tau)| ,
 \ee
which is valid everywhere except the points close to zeroes of $\sin (H\tau)$. In  the approximation  (\ref{a+}) we neglected the terms $\sim  \epsilon$ in accordance with  (\ref{hierarchy}). In particular,  $a(\tau=0)$ is in fact $\sim \sqrt{\epsilon}$ rather than  zero, and the exact solution (\ref{a1}) is required for the computation of $\eta$, see (\ref{eta2}).

Now we consider only single-folded instantons and compute the full period of conformal time $\eta$  which can be rewritten as follows
    \be
   \label{eta1}
     \eta= \frac1H\left[\int^{\pi/2}_0\frac{d\phi}{a(\tau)} +\int^{\pi}_{\pi/2}\frac{d\phi}{a(\tau)} \right],
     ~\phi\equiv 2H\tau,~~~
     \ee
and reduced to incomplete elliptic integral. Within the $\ln\epsilon$ accuracy it reads
       \be
   \label{eta2}
   \eta\simeq \frac{1}{\sqrt{2}}\ln\frac{1}{\epsilon}.
   \ee
   During this long evolution represented by conformal Euclidean time (\ref{eta2}) the scale factor $a(\tau)$ makes some drastic changes in size as one can see from the following estimation
   \be
   \label{a2}
   \frac{a_+}{a_-}\simeq \frac1{\sqrt{CH^2}}
   \sim \frac{1}{\sqrt{\epsilon}}\gg 1.
   \ee
One should observe here that there is a qualitative difference with discussions of the Model-1 when the ratio (\ref{a2}) was always parametrically of order one. In the present Model-2 this ratio (\ref{a2}) could be parametrically very large which implies that the largest and smallest sizes in the garland construction could have parametrically  different scales.\footnote{Note that a large value of the ratio $a_+/a_-$ does not essentially affect the thermal history of the inflation in the Lorentzian spacetime modulo the determination of its original energy scale, because the low temperature primordial radiation in Eq.(\ref{AAA}) gets quickly diluted during inflationary expansion and does not contribute to the reheating at the exit from the inflationary scenario.}

We conclude with the following comment. Merely the existence of the radiation enforces us to consider  the topology $\mathbb{S}^3\times \mathbb{S}^1$. If one ignores the radiation and the presence of $\mathbb{S}^1$ then the system defined on $\mathbb{S}^3\times \mathbb{S}^1$ becomes defined on $\mathbb{S}^4$, in which case the corresponding contribution to the path integral  is strongly suppressed as argued in \cite{Barvinsky:2012qm}. Technically, this suppression occurs as a result of  the conformal anomaly which changes the sign of the classical  Euclidean action. In addition, the positive action which is generated due to the conformal anomaly is divergent at $a\rightarrow 0$ for   $\mathbb{S}^4$.
This divergence  leads to the infinitely strong suppression of these  vacuum  $\mathbb{S}^4$ configurations, see \cite{Barvinsky:2012qm} for the comments and details. One should also add that in the Model-2 the relevant $\mathbb{S}^1$ structure might be generated not only by radiation but also by the quantum interactions in strongly coupled gauge theories as argued in Appendix \ref{inst-quark} such that size of the $\mathbb{S}^1$ is a free parameter of the model
and it is determined by the dimensional parameter $\Lbar$ of the strongly coupled $\qcd$ gauge theory.

\section{How the holonomy inflation ends }\label{inflation_end}

The main goal of this section is to argue that the holonomy  inflation paradigm advocated in this work  is consistent with all presently available observations.  One should emphasize that a theory describing    the end of inflation  (similar to pre-reheating and reheating stages in conventional inflationary scenario) in our framework is yet to be developed. The required technique which would answer the relevant questions
are formulated   in subsection \ref{anomaly} by  items 1-4.  Therefore, this section should  be treated  as a description of a vision and foresight
for a future development rather than a final formulation of the theory describing    the end of inflation.

 We focus on three  items to demonstrate the consistency of the framework. First of all we want to argue that the  equation of state (EoS) almost identically coincides with the EoS which is normally attributed to the cosmological constant. Secondly, we want to argue that the ``non-dispersive"
vacuum energy  which plays the key role in this framework   is capable to transfer its energy to the real propagating gauge fields of the Standard Model (SM). Therefore, the topological   inflation    could end with a successful  ``reheating epoch". Finally, we  estimate the number of $e$-folds $N_{\rm infl}$ for this  framework to show  that it is perfectly consistent with presently available observations. 

\subsection{Equation of State}\label{sect:EoS}
We start with the following  generic remark. Consider the  holonomy which assumes a nontrivial value along  $\mathbb{S}^1$   directed in  time direction
as discussed in previous section. In this case the Hubble constant and the energy density
 remain constant even after the nucleation from the gravitational instanton in spite of the fact that   the topology of the manifold  is not  $\mathbb{S}^3\times \mathbb{S}^1$ anymore. Further to this point, the system     is not described by the Euclidean metric after the nucleation, but rather assumes the conventional Lorentzian signature.

The corresponding Hubble constant $H$ is unambiguously determined by the  dimensional parameter $\Lbar$ of a strongly coupled gauge theory
as equation (\ref{Hubble}) states. This solution after the nucleation corresponds to the inflationary (almost) de Sitter behaviour such that the  EoS  and parameter ${\rm a}(t)$ assume the form:
\be	
\label{infl-EoS}
  w \equiv \frac{P}{\rho}\simeq -1  , ~~  {\rm a}(t)\sim \exp (H t),
\ee
in accordance with equations (\ref{EoS}) and (\ref{omega}).

The inflationary regime described by  (\ref{infl-EoS}) would be the final destination of our Universe if the interaction of the $\qcd$ fields with SM particles were always switched off. One should emphasize that the driving force for this inflationary deSitter behaviour (\ref{infl-EoS}) in the Lorentzian space is not a local inflaton field $\Phi(x)$ which is not present in our system at all. Rather the driving
force should be thought as a Casimir type vacuum energy which is generated by numerous tunnelling transitions in a strongly coupled gauge theory determined by the  dimensional parameter $\Lbar$. Precisely this parameter replaces the dimensional parameters
from inflaton potential $V[\Phi(x)]$ which  cosmology practitioners normally use in their studies.

When the coupling of the $\qcd$ fields with SM particles  is switched back on, the end of inflation is triggered precisely by this interaction which itself is unambiguously fixed by the triangle anomaly as we discuss below.

\subsection{Anomalous coupling of the ``non-dispersive" vacuum energy with gauge fields }\label{anomaly}

Before we explain the structure of the relevant interaction we want to make few comments in order to explain the physical nature of such unusual coupling
between propagating and non-propagating degrees of freedom.
First of all, we have to remind that  the physics responsible for the generating of the ``non-dispersive" vacuum energy (dubbed  as a ``strange energy'' in \cite{Zhitnitsky:2013pna,Zhitnitsky:2014aja,Zhitnitsky:2015dia})  which eventually leads to the de Sitter behaviour  (\ref{infl-EoS})   can not be formulated in terms of any physical propagating degrees of freedom as discussed    in great details in section  \ref{interpretation}. Instead, the generation of this energy  can be explained in terms of tunnelling transitions between topologically distinct but physically identical $|k\ra$ states.

The corresponding technique  to describe these tunnelling transitions    is  normally  formulated in terms of the Euclidean path integral and the corresponding   field configurations  interpolating  between  distinct  topologically sectors. In conventional  QFT computations the corresponding procedure selects a specific superposition of the  $|k\ra$ states which  generates the $|\theta\ra$ state with energy $E_{\rm vac}(\theta)$. In the context of inflation,   when the background assumes a non-trivial geometry (instead of $\mathbb{R}^4$ in conventional case)  the corresponding computations become profoundly  more complicated, though the corresponding procedure is well defined:\\
1. One should describe  the  relevant Euclidean configurations satisfying the proper boundary conditions for a nontrivial geometry (similar to calorons with nontrivial holonomy,  reviewed  in Appendix \ref{sec:holonomy});\\
 2. One should compute the corresponding path integral which includes all possible positions and   orientations of the relevant gauge configurations;\\
 3. The corresponding computations for the vacuum energy $\rho$ and pressure $P$ must be done with all fields which couple  to $\qcd$ gauge theory. Precisely this coupling is responsible for transferring the vacuum energy to SM particles;\\
 4. As the last step, one should subtract the corresponding expression computed on $\mathbb{R}^4$ as explained in section  \ref{holonomy}. Precisely this remaining part of the vacuum energy is interpreted as the relevant energy which enters the Friedmann equation,
 and which cannot be removed by any subtraction procedure and cannot be renormalized by any UV counter terms.
 The corresponding  formulae for $\rho, P$ will depend, in general, on properties of the manifold and relevant coupling constants.

 While these steps are well defined in principle, it is not   feasible to  perform  the corresponding  computations because even the
 first step in this direction,   a finding the relevant Euclidean configurations satisfying the proper boundary conditions for a nontrivial geometry, is yet unknown.  Nevertheless, this procedure, in principle, shows that  the deSitter behaviour (\ref{infl-EoS})
 in this framework emerges without any local inflaton field $\Phi(x)$ as explained  in previous section \ref{sect:EoS} because the physical  force driving the inflation  has  completely different nature in this proposal.

Fortunately, the key ingredients  which are relevant for our future studies can be understood  in alternative way, in terms of the  auxiliary topological non-propagating  fields  $b(x, H)$ which effectively describes the relevant infrared physics (IR) representing the key elements of the steps 1-4 highlighted above.

The corresponding  formal technique is widely used in particle physics and condensed matter (CM) communities. For the convenience  of the readers we provide   (within our cosmological context)  the main ideas and  results  of this approach  in Appendix  \ref{BF-section}.   In particular, this approach is extremely useful in description of the topologically ordered phases when the IR physics is formulated in terms of the  topological Chern-Simons (CS) like Lagrangian. One  should emphasize that the corresponding
  physics, such as calculation  of the braiding phases between quasiparticles, computation of the degeneracy etc,  can be computed (and in fact originally had been computed)  without Chern-Simons Lagrangian and without auxiliary fields. Nevertheless, the discussions of the IR physics in terms of CS like effective action  is proven to be very useful, beautiful and beneficial. In our case,
unfortunately, we cannot proceed with explicit computations along the  lines 1-4 as explained above.
  Therefore, the alternative technique in terms of the auxiliary topological non-propagating fields is the only remaining  option in our case.

  We refer to Appendix \ref{BF-section} where we overview the corresponding technique    in context of the  inflationary cosmology.
  We also explain there the physical meaning of these auxiliary field $b(x, H)$ which should be thought as the source of the topological
  fluctuations, similar to the axion field, see below.
Precisely this auxiliary non-propagating field      eventually generates the ``non-dispersive energy''  (\ref{Delta}) and  consequently leads to the de Sitter behaviour (\ref{infl-EoS}). This auxiliary field $b(x, H)$  effectively  describes (through the correlation functions) the modification  of the tunnelling rates   between topological  $|k\ra$ sectors as result  of external background field parameterized by   $H$. In other words,  a profoundly  complicated procedure of summation over all topological configurations interpolating between $|k\ra$-sectors  in the background parametrized by  $H$ as outlined  above (steps 1-4), can be expressed  in terms of the auxiliary field $b(x, H)$ which, of course,  remains  the   non-propagating auxiliary field in background $H$.

The only information which is required  for the future analysis    is that the relevant auxiliary field $b(x, H)$, saturating the ``non-dispersive" vacuum  energy (\ref{Delta}), couples to the SM particles precisely in the same way as the   $\theta$ parameter couples to the gauge fields.
This   claim is explained in Appendix \ref{BF-section}  and is based on analysis of the exact anomalous Ward Identities.
In many respects  the coupling of the  $b(x, H)$ field to the gauge fields is unambiguously  determined similar to unique coupling of the $\eta'$ field to the gluons,  photons and gauge bosons.

As a consequence of this fundamental feature  the topological auxiliary   $b(x, H)$ field is in fact an angular topological variable and it   has the same $2\pi$ periodic properties as the original $\theta$ parameter.
As it is known the $\theta$  parameter can be promoted to the dynamical axion field $\theta(x)$ by addicting the canonical kinetic term
$[\partial_{\mu}\theta(x)]^2$ to the effective Lagrangian. The difference of  the $b(x, H)$ field  with the dynamical axion $\theta(x)$ field is that the auxiliary topological field $b(x, H)$ does not have a conventional axion kinetic term.

For simplicity we also assume that $\qcd$ has a single flavour $N_f=1$ quark which couples to the  non-abelian $\qcd$ gauge gluons as well as to the $E\&W$ gauge fields, similar to conventional QCD quarks.
This is precisely the coupling which provides the interaction between the (conjectured) high energy $\qcd$ and the low energy $E\&W$ gauge fields.
It is naturally to assume that the mass of the corresponding $\bar{\eta}'$ is of order $m_{\bar{\eta}'}\sim\Lbar$, similar to the QCD case. Therefore, this heavy degree of freedom can be safely ignored in what follows.
In other words, the desired coupling of $b(x, H)$ auxiliary field with $E\&M$ gauge field is \cite{Zhitnitsky:2013pna}
\be	\label{coup}
  {\cal L}_{b\gamma\gamma} (x)= \frac{\alpha (H)}{8\pi} N  Q^2 \left[ \theta- b(x, H)\right] \cdot F_{\mu\nu} \tilde F^{\mu\nu} (x) \, ,
\ee
where $\alpha(H)$ is the fine-structure constant measured during the period of inflation, $Q$ is the electric charge of a $\qcd$ quark, $N$ is the number of colours of the strongly coupled $\qcd$, and $F_{\mu\nu}$ is the usual electromagnetic field strength. As we already mentioned,
the coupling (\ref{coup}) is unambiguously fixed because the auxiliary $b(x)$ field always accompanies the so-called $\theta$ parameter in the specific combination $\left[\theta-b(x, H)\right]$ as explained in  Appendix \ref{BF-section}, and describes the anomalous interaction of the topological auxiliary $b(x, H)$ field with $E\&M$ photons. In formula (\ref{coup}) we also ignored the heavy $\bar{\eta}'$ field which couples in the same way as auxiliary $b(x, H)$ field, i.e. $ \left[ \theta- \bar{\eta}'-b(x, H)\right]$. However, $ \bar{\eta}'$ field is very heavy as explained above, in contrast with   auxiliary  field which generates a topologically protected pole as explained in Appendix \ref{BF-section}.

The coupling  of the $b(x, H)$ with other $E\&W$ gauge bosons can be unambiguously reconstructed as explained in \cite{Zhitnitsky:2013pna}, but we keep a single $E\&M$ field $F_{\mu\nu}$ to simplify the notations and emphasize on the crucial elements of the dynamics, related to the helical instability which triggers the end of inflation, see next subsection \ref{instability}.

Based on coupling (\ref{coup}) we present our  numerical estimates for number $N_{\text{infl}}$ of e-folding in section \ref{e-folding}. 
Finally, in   subsection \ref{interpretation:emission} we interpret   the obtained results  and give an intuitive   explanation  why and how the non-dynamical auxiliary field $b(x, H)$ can, nevertheless,   produce  the real physical propagating degrees of freedom in a time-dependent background parametrized by $H$.

\subsection{The helical instability and the end of inflation }\label{instability}
It has been known for quite sometime that the structure of the interaction (\ref{coup}) in many respects has a unique and mathematically beautiful structure with a large number of very interesting features.
The most profound property which is crucial for our present analysis of the inflationary Universe is the observation that the topological term (\ref{coup}) along with the conventional Maxwell term $F_{\mu\nu}^2$ leads to an instability with respect to photon production if   $\dot{b}(x,H)$  does not vanish.
This is the so-called helical instability and has been studied in condensed matter literature \cite{Frohlich:2002fg} as well as in particle physics literature including some cosmological applications \cite{Joyce:1997uy}.

In context of our studies, the closest system where the helical instability develops is the system of heavy ion collisions \cite{Akamatsu:2013pjd} wherein $\la \dot{b}(x,H)\ra$ can be identified\footnote{The simplest way to demonstrate the correctness of this identification is to perform the path integral $U(1)_A$ chiral time-dependent transformation to rotate away the coupling (\ref{coup}).
The corresponding interaction reapers in the form of a non-vanishing axial chemical potential $\mu_5$, see Appendix B of ref.\cite{Zhitnitsky:2013pna} with details and references.} with the so-called axial chemical potential $\mu_5$.
One can explicitly demonstrate that the interaction (\ref{coup}) leads to the exponential growth of the low-energy modes with
\be
\label{mu_5}
k\leq \frac{\alpha (H) \mu_5}{\pi}, ~~~\mu_5\equiv \la \dot{b}(x,H)\ra .
\ee
The growth (\ref{mu_5}) signals that the instability of the system with respect to production of the real photons develops \cite{Akamatsu:2013pjd}.
It is also known that the fate of this instability is to reduce the axial chemical potential $\mu_5$ which was the source of this instability.
In  the inflationary context the corresponding instability reduces $H$ which plays the role of $\mu_5$, see discussions  below.
One should also comment here that parameter $\mu_5$ in heavy ion collisions is also not a dynamical field, but rather is an auxiliary fluctuating field which accounts for the dynamics of the topological sectors in QCD, similar to our case when $\la \dot{b}(x,H)\ra$ describes the dynamics of the topological sectors in $\qcd$.

This short detour into the nature of helical instability as a result of interaction (\ref{coup}) has direct relevance to our studies because the auxiliary field $b(x, H)$ entering eq.(\ref{coup}) exhibits all the features of parameter $\mu_5$ which was the crucial element in the analysis of the helical instability in heavy ion collisions.
Indeed, both these auxiliary fields originated from the same physics and they both describe the dynamics of the topological sectors  in strongly coupled gauge theories.

In  terms of physics these non-propagating fields  effectively account for the long range variation of the tunnelling processes as a result of some external influence of the backgrounds expressed in terms of $H$ for inflation and in terms of $\mu_5$ for heavy ion collisions respectively,
see some additional comments on this analogy in  Appendix B of ref.\cite{Zhitnitsky:2013pna}.

The net result of the interaction (\ref{coup}) and instability (\ref{mu_5}) is that the holonomy inflation in this framework inevitably  ends by transferring the ``non-dispersive" vacuum energy proportional to $H$ as eq.  (\ref{Delta2}) states into the real   propagating gauge fields. One can interpret this energy transfer as a back-reaction to the auxiliary field $b(x, H)$ as a result of adjustment of the system due to the interaction (\ref{coup}). How this back-reaction effect can be in principle computed? The corresponding computations based on  the first principles as listed in Section \ref{anomaly} by items 1-4 are not presently feasible as we already mentioned. Effective description in terms of the dynamics of the  auxiliary field $b(x, H)$ can be, in principle, carried out along the line mentioned at the very end of Appendix \ref{BF-section}. 

One may also wonder if entire vacuum energy will be transferred to the radiation in the form of
 the SM gauge field, which is the key element for successful
graceful exist from inflation. Our comment here is that the transfer of the vacuum energy in this framework is a continuous process, rather than a one-time event.    This is obviously the same back-reaction effect which is mentioned in previous paragraph: the radiation decreases the magnitude of the vacuum energy.  This process continues as long as the vacuum energy still remains a source of the radiation.
This process lasts as long as $ \la \dot{b}(x,H)\ra\neq 0$.

The physical picture of this energy transfer is as follows\footnote{The  intuitive picture presented below is based on our understanding of the
fate of the helical  instability 
in heavy ions collisions leading to reduction  the axial chemical potential $\mu_5$ which itself is the source of this instability.}.  Non-vanishing value for $ \la \dot{b}(x,H)\ra\neq 0$ leads to the particle production. This radiation of particles obviously decreases the value of
 $ \la \dot{b}(x,H)\ra$ (and the corresponding vacuum energy) as the source of this radiation. In terms of real  physical processes this  energy transfer  corresponds to
the modification of the tunnelling transition rate with emission of the real particles in a nontrivial background which also varies. The radiation continues as long as the background deviates from the flat Minkowski space-time.

The technical description of this energy transfer cannot be carried out in conventional way, let us say,  in terms of physical propagating degree of freedom. For example,  we cannot  model these radiation processes by adding a kinetic term to  ${b}(x,H)$ field  because the corresponding anomalous Ward Identities cannot be satisfied with physical propagating degrees of freedom as explained in Appendix \ref{BF-section}. 
 We think, that 
he holographic description  mentioned in  Appendix \ref{BF-section} offers a possible framework which potentially can
accommodate the dynamics of the auxiliary $b(x, H)$ field, strange features of the non-dispersive vacuum energy and back-reaction effects due to the coupling with the SM fields (\ref{coup}). At present time we do not know yet how to   formulate a proper    computational framework to answer this question.

To conclude this subsection we would like to comment here that the energy transfer between non-dynamical auxiliary fields and propagating dynamical fields can be in principle tested in a tabletop experiment based on  the Maxwell system.
   We explain the relevant physics   and also offer a possible design for a tabletop experiment   in  subsection \ref{interpretation:emission} where such unusual  effect can be, in principle,  experimentally tested in a simplified settings.

  \subsection{Estimates for the e-folds}\label{e-folding}
The number of $e$-folds in the holonomy inflation is determined by the time $\tau_{\rm inst}$ when the helical instability fully develops, which explains
our subscript $\tau_{\rm inst}$.  This is exactly the time scale where a large portion of the energy density $\rho$  from eq. (\ref{Delta2}) which eventually generates the Hubble constant $H$ according to (\ref{Hubble}) is transferred to SM light fields.
The corresponding time scale for the heavy ion system is known \cite{Akamatsu:2013pjd} and it is given by $\tau_{\rm inst}^{-1} \sim \mu_5\alpha^2$.
For our system $\mu_5$ should be interpreted as  $\la \dot{b}(x, H)\ra\sim H$, as the only relevant scale of the problem, see also few additional arguments in Appendix \ref{BF-section} supporting this estimate. At the moment $\tau_{\rm inst}$  the de Sitter
growth  (\ref{infl-EoS}) cannot be maintained anymore as the source of this behaviour $\sim H$  is completely exhausted  due to the    transferring  its energy to the gauge fields of the SM.

Therefore we arrive at the following order of magnitude estimate for the number of $e$-folds $N_{\text{infl}}$ in $\qcd$ inflationary paradigm,
\be	\label{e-folds}
  \tau_{\rm inst}^{-1}\sim  {H\alpha^2(H)}, ~~~~\Longrightarrow ~~~N_{\text{infl}}\sim \frac{1}{\alpha^2(H)} \sim 10^2,~~~~
\ee
where number of $e$-folds $N_{\text{Inf}}$ is, by definition, the coefficient in front of $H^{-1}$ in the expression for the time scale $\tau_{\rm inst}$.
At this moment the  energy density $\rho$  from eq. (\ref{Delta2})  ceases to exist as the dominant portion of the energy of the system.

The key element of this holonomy inflationary scenario is that the number of $e$-folds $N_{\text{infl}}$ when  the de Sitter behaviour (\ref{infl-EoS}) ends is determined  in this framework by the gauge coupling constant $\alpha(H)$ rather than by dynamics of ad hoc inflaton filed $\Phi$ governed by some ad hoc inflationary potential $V(\Phi)$.

\exclude{
\subsection{ Holonomy inflation confronting the Observations}\label{observations}
The goal of this subsection is to present a possible computation technique to study the cosmological perturbations in this framework. One should emphasize that the corresponding computation technique is not developed yet, and there are many remaining obstacles to produce some solid results.  While the basic steps (1-4)  of the  computations (including the analysis of  the cosmological perturbations) as highlighted in section \ref{anomaly} in principle,  is a well defined procedure, the practical  applications are hard to carry out. We explain some of the technical obstacles
to carry out such computations  in the following section \ref{interpretation:emission}.

If we  could carry out the  four steps as listed in section \ref{anomaly} we  would be able to compute the  equation of state $w$, pressure $P$, and the vacuum energy $\rho$ as a function of time. Precisely this information enters the formulae for the fluctuations as we discuss below.

Normally,  the computations of the gravitational potential $\delta_{\Phi}^2$ and  the tensor fluctuations $\delta^2_h$ are  carried out using  specific features of the inflaton field $\phi$ and the scalar potential $V[\phi]$. In our framework we do not have a dynamical scalar field $\phi$.
However, it is naturally to expect that the source for both, the power spectrum $\delta_{\Phi}^2$ and  the tensor fluctuations $\delta^2_h$
 must be expressed in terms of the energy-momentum tensor as  the only gauge invariant characteristic of the system, irrespectively   whether  the energy-momentum tensor is expressed in terms of scalar inflaton, or any other fields, such as vector fields, gauge fields, or vacuum gauge configurations describing the tunnelling events.

 The corresponding formulae for  the power spectrum $\delta_{\Phi}^2$ and  the tensor fluctuations $\delta^2_h$ in terms of the
  pressure $P $ and vacuum energy $\rho$ have been derived in \cite{mukhanov}, and we shall use the corresponding expression in our estimates which follow. In other words, we assume that corresponding expressions for fluctuations in terms of the energy-momentum tensor  must remain the same  irrespectively  the nature of the source of the pressure $P$, and the vacuum energy $\rho$.
  In principle,  to justify this assumption one should  derive the fluctuations using the procedure highlighted in section \ref{anomaly} as steps 1-4.
  As we already mentioned we do not have the  technical  tools to proceed with such computations. Therefore, we proceed with the
   assumption that the formulae derived \cite{mukhanov} using conventional scalar inflaton potential still holds even when sources
   have different nature, not expressible in terms of the local scalar field $\phi$.

Indeed, the power spectrum in the post-inflationary, radiation dominated epoch can be expressed in terms of these parameters as follows \cite{mukhanov}:
\be
\label{power}
\delta_{\Phi}^2\simeq \frac{64}{81}\left(\frac{\rho}{|c_s|(1+P/\rho)}\right)_{c_sk\simeq Ha}
\ee
where we use notations from textbook \cite{mukhanov}. In this formula   the so-called speed of sound\footnote{One should comment here that in the conventional description with inflation described in terms of the physical propagating degrees of freedom the parameter $c_s^2$ must be positive.
In such a conventional case a negative $c^2_s<0$ is considered as a signal of instability of the system.
There is no such requirement for our system as there are no any propagating degrees of freedom associated with this speed $c_s$.
In particular, in a pure de Sitter state $c_s^2=-1$ as one can see from (\ref{infl-EoS}), and it is obviously consistent with all fundamental theorems, see also an additional comment on $c_s$ in footnote 4 in ref.\cite{Zhitnitsky:2013pna}.}
can be also expressed in terms of pressure $P$, and the vacuum energy $\rho$ as a function of time \cite{mukhanov}. In our system this parameter is not known, but in principle, is calculable from the first principles, similar to  computations of $P$ and $\rho$ as highlighted above. In what follows to simplify the estimates we assume $|c_s|=1$.

 Similarly, the expression for the ratio of tensor to scalar power spectrum amplitudes during the post-inflationary, radiation dominated epoch can be expressed in terms of the same  parameters as follows \cite{mukhanov}:
\be	\label{r}
  r\equiv \frac{\delta^2_h}{\delta_{\Phi}^2}\simeq 27 \left[|c_s|\left(\frac{P}{\rho}+1\right)\right]_{ k\simeq H{\rm a}},
\ee
while the tensor spectral index is given by \cite{mukhanov}
\be	\label{nt0}
  n_t\simeq -3 \left(\frac{P}{\rho}+1\right)_{k\simeq Ha}.
\ee
The spectral index $n_s$   is also known in terms of the EoS and it is given by \cite{mukhanov}
\be	\label{ns}
  n_s-1\simeq -3 \left(\frac{P}{\rho}+1\right)- \frac{1}{H}\frac{d \ln\left(\frac{P}{\rho}+1\right)}{dt}-\frac{1}{H}\frac{d \ln |c_s|}{dt}.~~~~~~
\ee

Therefore, our claim is that one can compute pressure $P$,  the vacuum energy $\rho$ from the first principles as a function of time as discussed in section  \ref{anomaly} to make connection with observations without referring to any auxiliary fields, or other additional elements of the framework. Technically, it is not feasible at the present time as we already mentioned.

Therefore, we would like to  introduce a number of additional technical elements and some  phenomenological parameters which effectively account for the particle production resulted from  the helical instability. The corresponding physics is very different from conventional and well known computational technique when a dynamical inflaton field $\Phi$ couples to other physical fields and transfers the energy as a results of this interaction. In our case the coupling is between SM gauge fields and auxiliary non propagating field. Therefore, the corresponding energy transfer is formulated very differently from conventional framework.  We explain some conceptual steps for this computation scheme (which is yet to be developed) in section \ref{interpretation:emission}.

Now we follow a simplified procedure suggested in  \cite{Zhitnitsky:2014aja}
and  parametrize the physics  at the very end of  inflation
resulted from  the helical instability  as follows \cite{Zhitnitsky:2014aja}:
\be	
\label{eos}
  w=\frac{P}{\rho} = -1+ c_2\alpha^2(H_0)\cdot e^{c_1\left[\frac{t-t_i}{\tau_{\rm inst}}-1\right]},
\ee
where the two numerical coefficients $c_1, c_2$ will be fixed bellow using two  observables: $n_s-1$ and $r$.
The parametrization (\ref{eos}) properly accounts for the exponential growth of the produced particles  due to the helical instability, see explanations  in \cite{Zhitnitsky:2014aja}.

We fix these two constants by fixing $w$ and its time derivative during the final moment of inflation as follows:
\be	\label{const}
  \left(\frac{P}{\rho}+1\right)_{t-t_i=\tau_{\rm inst}}=c_2\alpha^2(H_0)\\
  \frac{d \ln\left(\frac{P}{\rho}+1\right)}{dt}|_{t-t_i=\tau_{\rm inst}}=\frac{c_1}{\tau_{\rm inst}}.\nonumber
\ee
where $\tau_{\rm inst}$ is the time when the helical  instability develops which triggers the end of the inflation in our framework.

 We are now in position to fix the two free parameters from (\ref{eos}) using the measured values for the spectral index $n_s\simeq 0.96$.
 The   tensor fraction $r<0.12$  has not been measured yet. For illustrative purposes we assume $r\simeq 0.05$ in our estimates below.
 The corresponding estimates can be always updated when more precise constraint (or observation) is available.

We start our analysis  with $r\simeq 0.05$.
In conventional inflationary scenarios based on the scalar potential the magnitude of $r$ is normally expressed in terms of the slow-roll parameters of the inflaton potential $V(\Phi)$.
In our framework we do not have scalar field, nor scalar potential $V(\Phi)$. Nevertheless, the EoS is perfectly defined for the system. In fact, all observables can be directly expressed in terms of the EoS without even mentioning the potential $V(\Phi)$.
In particular, the expression for tensor fraction $r$ is given by (\ref{r}).
We assume $|c_s|^2\simeq 1$  to  arrive to the following  condition which determines our parameter $c_2$,
\be	\label{r1}
  r\simeq 27  c_2\alpha^2(H_0)\simeq 0.05.
\ee
Similarly, one can estimate parameter $c_1$ using expression (\ref{ns}) for $(n_s-1)$
\be	\label{ns1}
  n_s-1 &\simeq&  -3c_2\alpha^2(H_0)-\frac{c_1}{H_0\tau_{\rm inst}} \nonumber\\
        &\simeq& -\alpha^2(H_0) \left[3c_2+c_1\right]\simeq -0.04,
\ee
where we used $n_s\simeq 0.96$.
Assuming $N_{\rm infl} \simeq 100$ and using estimates (\ref{e-folds}), (\ref{r1}), and (\ref{ns1}), we find the following set of parameters which approximately describe the observations:
\be	\label{fit}
  c_2\simeq 0.18, ~ c_1\simeq 3.44, ~ \alpha(H_0)\simeq 0.1, ~N_{\rm infl} \simeq 100.
\ee
We want to emphasize that it was
  not our goal to fit the data with perfect accuracy.
Such an analysis would be too premature at this point as a numerical understanding of the evolution of the helical instability (which determines the EoS) is yet to be fully developed. We do not know the parameter $r$ at this point, so our guess (\ref{r1}) could be far away from the reality.
Rather, our goal was to demonstrate that the holonomy inflation, in principle, can easily accommodate the presently available observations without any fine-tuning adjustments.

 One should also mention that the power spectrum for the scalar perturbations  (\ref{power}) is also unambiguously expressed in terms of
 $P, \rho$ parameters. The expression  (\ref{power})  can be  easily  fitted to be consistent with observations because the Hubble parameter $H$  is small in Planck units and expressed in terms of $\Lbar$ according to (\ref{Hubble}). We cannot unambiguously fix the scale
 $\Lbar$ in terms of $H$ at this point because the energy density $\rho$ in our computations also depends on dimensionless  parameter $\bar{c}_{{\cal{T}}}$ which, in principle, calculable from the first principles, but is yet unknown, as discussed in section \ref{interpretation}.

Now we are in a position to make some predictions for observables which have not been measured yet.
As an example, let us consider the tensor tilt $n_t$.
The corresponding expression in terms of the EoS is known
\be	\label{nt}
  n_t\simeq -3 \left(\frac{p}{\rho}+1\right)\simeq -3\alpha^2(H_0)c_2\simeq -5\cdot 10^{-3}.
\ee
where for a numerical estimate we use (\ref{const}) and (\ref{fit}).
It is interesting to note that our estimate is consistent with conventional predictions of slow roll inflation where $n_t\simeq -r/8$.
}
In next subsection \ref{interpretation:emission} we explain the concept of mechanism of the energy transfer at the end of inflation.  It is very different from conventional mechanism when propagating inflaton $\Phi$ couples with physical particles and transfer the energy.   In subsection \ref{relation} we compare our framework with conventional inflationary scenario to show some similarities and differences between the two approaches.

\subsection{Interpretation}\label{interpretation:emission}
In this subsection we want to explain a fundamentally new type of particle  production which is the key element in
all our discussions in this section related to the question how the inflation ends in this framework due to the coupling (\ref{coup}) of the auxiliary field with real physical gauge fields from the SM.

The main point is that the driving force for inflation in this framework is the non-dispersive vacuum energy which generates the EoS given by (\ref{infl-EoS}). Without anomalous coupling (\ref{coup}) it would be the final destination of the Universe. How does this coupling produce the particles? The main point is that the topological fluctuations with the typical scale $\sim \Lbar$  which saturate the vacuum energy are slightly different in the presence of background with scale $\sim H\ll \Lbar$.  This time dependent background generates the particle production with the rate $\sim H$ which is precisely  the reason why inflation eventually ends in this framework on the time scale (\ref{e-folds}).

We want to test  this mechanism  of the particle production  from ``non-dispersive" vacuum energy using the Maxwell theory as a playground.
\exclude{The effect of particle production  from ``non-dispersive" vacuum energy (which is the key element of this section)    is highly nontrivial novel phenomenon which obviously deserves
further studies and analysis.  In this subsection we would like to explain the corresponding effect and develop some intuitive picture  by presenting a toy model for this phenomenon using the Maxwell theory which is the part of the SM. }The corresponding Maxwell system can be, in principle, designed and fabricated
with existing technology, see the relevant references in Concluding Section \ref{test}. Therefore, in principle,   this novel phenomenon can be  tested in a  tabletop experiment in  a lab.

The basic idea is that there is
a  new contribution to the Casimir pressure which emerges as a result of tunnelling processes when the Maxwell system is formulated on a nontrivial manifold   permitting   the $E\&M$ configurations with nontrivial topology $\pi_1[U(1)]=\mathbb{Z}$. Precisely these tunnelling transitions between physically identical but topologically distinct states play the same role in the Maxwell system as the topologically nontrivial configurations in $\qcd$.
The corresponding extra energy generated due to these transitions is the direct analog of the ``non-dispersive" contribution to the energy     which is the key player of the present work as it explicitly enters (\ref{Delta2}), (\ref{consistency}), (\ref{consistency1}) in previous sections.
This ``non-dispersive" energy in the Maxwell system  is  similar to our studies  of the  non-abelian  gauge theories  reviewed in Appendix \ref{review} this extra energy also  cannot be formulated in terms of conventional propagating photons with two transverse polarizations.

If the same system is considered in the background of a small external time-dependent   field, then real physical particles  will be emitted from the vacuum, similar to the dynamical Casimir effect (DCE) when photons are radiated from the vacuum due to time-dependent boundary conditions.
Essentially, the ``reheating epoch" as advocated in this section when the vacuum energy can radiate real particles in a time dependent background
is analogous to the DCE.  The difference is that in  conventional DCE  the virtual particles from vacuum become real propagating particles in a time dependent background and get emitted. In our case the $E\& M$ configurations which describe the  interpolations  between different topological sectors
get excited in time dependent background and emit real particles, see concluding section \ref{test} for references and details.

We hope this intuitive explanation  provides the basic  conceptual picture   on how the particles can be produced from the vacuum, which represents  the key element of the graceful exit from inflation.

\subsection{Relation to the conventional inflationary scenario }\label{relation}
The goal of this section is to collect a number of comments made in different places in this work
related   to the  (possible) connection  between our framework and conventional description in terms of a scalar inflation $\Phi(x)$ governed by a potential $V[\Phi]$. By obvious reasons this is not a one to one correspondence
between drastically different descriptions. Nevertheless, these comments, hopefully,  may generate some thoughts
about the source  of the vacuum energy in Nature, and find a proper technical framework to describe it.

We start with few generic remarks.
The  topological  inflationary mechanism as formulated in this proposal is fundamentally non-local in nature and cannot be modelled by any local effective inflationary potential $V(\Phi)$.  Furthermore, this mechanism is fundamentally ``no-dispersive" in nature and cannot be described  in terms of  any propagating physical degree of freedom such as inflaton $\Phi(x)$ with   canonical kinetic term $\left(\partial_{\mu}\Phi(x)\right)^2$.
Further to this point. We introduced the topological auxiliary fields $a(x, H)$ and $b(x,H)$ in Appendix \ref{BF-section} to describe the physics in terms of effective long range fields which
in principle  should describe the  relevant  IR physics. These fields are not propagating, in contrast with the   inflaton $\Phi(x)$ field. The physical meaning of these fields as explained in Appendix  \ref{BF-section} is:  the $\Box a(x,H)$ describes the distribution of the topological density in the system, while  $b(x,H)$ acts as the axion field (without kinetic term) being the source of the topological density distribution.

These obvious differences between drastically different frameworks must obviously lead to distinct  observational  results. 
In particular, the conventional computations of the cosmological perturbations are based on treating the inflaton $\Phi(x)$ as the conventional scalar field with   canonical kinetic term $\left(\partial_{\mu}\Phi(x)\right)^2$. The corresponding results can be expressed in terms of the  
 vacuum energy $\rho$ and pressure $P$ as it is formulated in \cite{mukhanov}. However,  merely  existence of a local inflaton field $\Phi(x)$
 has been   assumed in computations in \cite{mukhanov}, while the final results are presented in terms of energy-momentum tensor. 
 Computations in our framework requires a different technique, which is not yet developed as explained at the very beginning of this section \ref{inflation_end}. Therefore, it is naturally to expect that the outcome would be different even when the final results are expressed in terms of the energy-momentum tensor's  parameters $\rho$ and $P$.   However, as the corresponding technical tools are  not yet developed, it is very hard to quantify the corresponding differences.  

In what follows we want to make few    comment on some similarities between these two distinct approaches. 
In particular,  we would like to  identify  (on intuitive level) the topological auxiliary fields $a(x, H)$ and $b(x,H)$ with the inflaton  $\Phi(x)$  field
in a sense that both fields eventually generate the deSitter behaviour, and both approaches lead to the inflationary EoS (\ref{infl-EoS}). The fundamental difference between the two is that the inflaton $\Phi(t)$ satisfies the classical equation of motion and depends on time $t$, while  $a(x, H)$ and $b(x,H)$ are truly quantum objects, such that all observables in principle must be 
 expressed exclusively in terms of the correlation functions and expectation values when the time dependence enters the physics  exclusively in terms of the Hubble parameter $H$.

Still, there are some hints which apparently suggest that some links between the two approaches may exist.

Indeed, let us introduce few important  parameters  which are normally used in conventional inflationary analysis and compare them with our description. For this purposes we introduce conventional slow-roll parameters, see e.g. \cite{linde}:
\begin{equation}
\label{slow-roll}
\epsilon=\frac{M_P^2}{2}\left(\frac{V_{\phi}}{V}\right)^2, ~~~
V_{\phi}\equiv\frac{\partial V[\phi]}{\partial\phi},
\end{equation}
\exclude{
In terms of these parameters the spectral indices are expressed as follows \cite{linde}:
\be
\label{indices}
n_s-1=-6\epsilon +2\eta, ~~~ r=16\epsilon, ~~~ n_t=-2\epsilon=-\frac{r}{8},~~
\ee
while the amplitudes for the perturbations    assume the form \cite{linde}:
\be
\label{amplitudes}
A_s= \frac{V}{24 \pi^2\epsilon} ~~~~~~~~ A_t=\frac{2V}{3\pi^2}.
\ee

In conventional framework these parameters  are numerically small as a result of properties of the inflaton potential (\ref{slow-roll}).
 
One can compare the conventional description (\ref{indices})  in terms of slow-roll parameters
with our analysis in section \ref{observations} to arrive to important conclusion that all these indices are numerically small in our framework because they all proportional to small coupling constant $\sim\alpha^2$. Indeed, $r, (n_s-1), n_t$ are $\sim \alpha^2$ according to (\ref{r1}), (\ref{ns1}) and (\ref{nt}) correspondingly. In conventional framework, on other hand,  these parameters are numerically small because they are proportional to specifically adjusted properties of the inflaton potential (\ref{slow-roll}).

  In the holonomy framework these amplitudes are parametrically  small because they are   proportional  to the
new scale $\Lbar^6/M_P^2$ according to (\ref{Hubble}). Precisely this  suppression provides the numerical smallness
of the amplitudes (\ref{amplitudes}) in the holonomy  framework.
 
It is interesting to note that even some parametrical (not numerical) enhancement  in $A_s$ in form of the $1/\epsilon$   in (\ref{amplitudes}) has its counterpart in the holonomy inflation  when $A_s$ has an additional enhancement factor $(1+\frac{P}{\rho})^{-1}\sim \alpha^{-2}$ according to (\ref{power}) and (\ref{const}). Therefore, it is not really a surprise that both approaches
predict a small  magnitude for $r$ according to (\ref{r}) and for  $n_t$ according to (\ref{nt}) as both are proportional to one and the same suppression factor $(1+\frac{P}{\rho})\sim \alpha^2$.
}
For example, the computation of the number of e-foldings   in conventional slow roll approximation and estimates
(\ref{e-folds}) in the holonomy inflationary scenario both produce numerically large magnitudes. 
In the conventional approach one can  use the following relation \cite{linde}:
\be
\label{N-folding}
N_{\rm infl}\simeq \frac1{M_P^2}\int^{\phi}_{\phi_{\rm end}}d\phi\left(\frac{V}{V_{\phi}}\right).
\ee
The large numerical value for $N_{\rm infl}\gg 1 $ in the conventional approach is due to the specific choice of the potential  (\ref{slow-roll}) when the integrand entering (\ref{N-folding}) is parametrically  large and proportional to $\epsilon^{-1}$. It should be compared with  the holonomy inflationary scenario when $N_{\rm infl}\gg 1 $ is parametrically large due to the  enhancement factor  $\alpha^{-2}$ as estimates (\ref{e-folds}) suggests.

We conclude this section with few generic comments.
 First of all, while we identify (on the intuitive level) the auxiliary  topological fields with inflaton, the $a(x, H)$ and $b(x, H)$ fields   remain to be quantum (not classical) fluctuating fields, saturating the relevant correlation functions. We observed above that there is a number of instances when the holonomy inflationary scenario behaves very much in the same way as the conventional  description represented by formulae (\ref{slow-roll}),  (\ref{N-folding}) discussed above. Is it  a coincidence or there is a more deep reason for these relations?

 We formulate the same question in a different way: Is it possible to make any connection between with conventional description in terms of auxiliary $a(x, H)$ and $b(x,H)$ fields and     local inflaton $\Phi(x)$ field  which satisfies the classical equation of motion
 determined by the potential $V[\Phi]$? We do not know how to do it. The main obstacle to make such a connection is related  to the fact that  the auxiliary topological fields, by construction (reviewed   in Appendix \ref{BF-section}) saturate the topological susceptibility (and the corresponding vacuum energy) with the positive sign according to (\ref{top1}) and (\ref{K1}), generating the topologically protected pole (\ref{K}), while any conventional degree of freedom (including dynamical propagating  inflaton) can only produce a negative sign according to (\ref{G}).

 One possible path to overcome this obstacle is to define the auxiliary fields\footnote{We remind the physical meaning of the auxiliary fields: the $\Box a(x,H)$ describes the distribution of the topological density in the system, while  $b(x,H)$ acts as the axion field (without kinetic term) being the source of the topological density distribution.} using the holographic description along the line suggested in \cite{Zhitnitsky:2011aa}. In this case the axion field which is represented by our auxiliary field $b(x,H)$ becomes the dynamical propagating field in the bulk of multidimensional space but acts as a conventional (non-dynamical) term on the boundary (representing our space-time). This feature is precisely what is required for  our auxiliary field  $b(x,H)$ defined on physical space-time.

\section{Concluding Comments}\label{conclusion}
We conclude this work with formulation of our basic results in section \ref{basic}.
We next formulate the profound consequences of our framework in section \ref {profound}.
To convince the readers that we study a real physical effect,  we suggest to  test  this new ``non-dispersive" type of  vacuum energy   in a laboratory
using the physical Maxwell system
as highlighted in section \ref{test}. Finally, we make few comments on relation of our approach with  no-boundary and tunnelling proposals in subsection \ref{no-boundary}.
\\
\subsection{Basic results}\label{basic}
The  heart of the proposal suggested  in the present work is a synthesis of two, naively unrelated,   ideas.

First idea represents the self-consistent treatment
of the problem    formulated on the Euclidean $\mathbb{S}^3\times \mathbb{S}^1$ manifold  through the bootstrap equation  \cite{Barvinsky:2006uh,Barvinsky:2006tu,why}.

The second novel idea   \cite{Zhitnitsky:2013pna,Zhitnitsky:2014aja,Zhitnitsky:2015dia} is a proposal to treat the vacuum energy entering the Friedmann equation as a ``non-dispersive"   vacuum energy which is  always generated in non-abelian gauge theories as  a result of tunnelling
transitions between topologically   nontrivial    sectors in  a system. This type of energy is very unusual in many respects.   First of all,   it is non-analytical  in coupling constant $\sim \exp(-1/g^2)$ and can not be seen in perturbation theory as reviewed in Appendix \ref{interpretation}.   Secondly,  this vacuum energy is non-local in nature as it cannot be expressed in terms of any local operators in a gradient expansion in any effective field theory. Rather, it can be expressed in terms of the non-local holonomy, similar to Aharonov-Casher effect as mentioned in section \ref{topology}.

We coin the marriage of these two sets of ideas as the holonomy inflation which has a number of very attractive and desirable features.
First of all, there is the  hierarchy of scales for both models given by eq. (\ref{hierarchy1}) and  (\ref{hierarchy2}) correspondingly which indicates that  the distances smaller than Planck scale $M_P^{-1}$ never appear  in our framework. Secondly, the Equation of State (\ref{infl-EoS})   assumes its de Sitter behaviour  as a result
of nucleation as Fig.\ref{Fig2} shows. Thirdly, the number of $e$-folds $N_{\rm infl}$ is naturally determined by the gauge coupling constant $\alpha (H)$ as equation (\ref{e-folds}) suggests. 

\subsection{Implications and future development}\label{profound}
There are few important and generic consequences of this framework.

1. The conventional  scenarios of the eternal self-producing inflationary universes are always formulated in terms of a physical scalar dynamical inflaton field  $\Phi(x)$. This problem with self-reproduction of the universe does not even emerge in our framework as there are no any fundamental scalar  dynamical  fields in the system responsible for inflation.
Instead, the   de Sitter behaviour   in  our framework  is pure quantum phenomenon, which     is a consequence of  the dynamics of the long ranged topological configurations with nontrivial holonomy, rather than a result of a physical fluctuating dynamical field.   This type of energy  manifests itself in terms of the ``wrong" sign in the correlation function which can not be formulated in terms of any local propagating degrees of freedom as explained in Appendix  \ref{contact}. Therefore, the problem with eternal inflation does not even occur in our framework.
\exclude{The corresponding  topological configurations which are responsible for this behaviour   may  generate, as argued in this work,  the linear in $H$ correction in the Friedman equation (\ref{Delta1}), (\ref{Delta2}) which eventually leads to the de Sitter behaviour as explained in this work.
}

2. There are many other problems in conventional formulation of the inflation in terms of scalar   inflaton field  $\Phi(x)$.
 For example,  the  initial value $\Phi_{\rm in}\gg M_{\rm PL}$ for the inflaton is normally very large. This problem does not occur in our holonomy inflation
 scenario as the hierarchy of scales   (\ref{hierarchy1}) always holds in our framework.

3. We should   also mention that  the energy described by a formula similar to eqs.(\ref{Delta1}), (\ref{Delta2}), which  eventually leads to the de Sitter behaviour (\ref{a}), has been previously postulated \cite{Urban:2009vy,Urban:2009yg,Urban:2009wb} as the driving force for the dark energy (admittedly, without much deep theoretical understanding behind the formula at that time). The model has been (successfully) confronted with observations\footnote{We note that the structure of the relevant vacuum energy which enters  the Friedman equation (\ref{Delta1}), (\ref{Delta2}) is determined by the size of $\mathbb{S}^1$ and behaves in all respects as the cosmological constant. Therefore, it  is obviously consistent with presently available data as it does not modify the equation of state as explained in Appendix \ref{interpretation}.}, see recent review papers \cite{Cai:2014pek,Cai:2012fq}   and many original references therein, where it has been claimed that this proposal is consistent with all presently available data.

Our comment here is that history of evolution of the universe may repeat itself by realizing the de Sitter behaviour twice in its history.  The $\qcd$-dynamics was responsible for the holonomy inflation considered in present work, while   the QCD dynamics is responsible  for the dark energy in present epoch. In this case the DE density is given by expression similar to (\ref{Delta2}), i.e. $\rho_{\rm DE}\sim H\Lambda_{\rm QCD}^3\sim (10^{-3} {\rm eV})^4$ is amazingly close to the observed value without any fine-tunings or adjustments of the parameters.

4. One should also mention that  some recent    lattice simulations  \cite{Yamamoto:2014vda}  implicitly support our results.
Indeed, the  author of ref.  \cite{Yamamoto:2014vda} studied the rate of particle production in the de Sitter background. The rate turns out to be linearly proportional to the Hubble constant $\sim H$, rather than naively expected $H^2$. It is fully consistent with our proposal\footnote{Indeed, the rate of the particle production in quantum field theory in general is determined by the imaginary part of the stress tensor, Im$ [T_{\mu}^{\nu}]$, while the vacuum energy is related to the real part of the stress tensor, Re$[T_{\mu}^{\nu}]$.  Analyticity suggests that both components must have the same corrections  on $H$ at small $H$. Therefore, the lattice measurements
 \cite{Yamamoto:2014vda} of the linear dependence on $H$ of the particle production strongly suggest that the vacuum energy (which is determined by the real part of the same stress tensor) must also exhibit the same linear $\sim H$ correction. The corresponding lattice computations of the $\theta$ dependent portion of the vacuum energy and topological susceptibility in time dependent background are possible in principle, but technically much more involved than the  analysis performed in ref. \cite{Yamamoto:2014vda}. }. We hope that some further lattice computations   in time dependent background can further elucidate  the role of holonomy in generating the   vacuum energy.

5. Finally,  we want to make a comment about possible future development. As we already mentioned at the beginning of Section \ref{inflation_end} the relevant technique describing the end of inflation in our framework  (including computations of the cosmological perturbations)   is yet to be developed.  
We already mentioned in the text a number of technical challenging problems  which need  to be resolved,  and shall not repeat them here in Conclusion. 

\subsection{Possible tests of the cosmological  ideas in a lab?}\label{test}
Our  comment here is that we cannot ``experimentally" test the first element of the proposal  advocated in \cite{Barvinsky:2006uh,Barvinsky:2006tu,why} in any simplified settings.
 However, we can test the second element of this proposal advocated in  \cite{Zhitnitsky:2013pna,Zhitnitsky:2014aja,Zhitnitsky:2015dia} in  tabletop experiments. This subsection should convince the readers that we are dealing with a   new physical phenomena which can be realized in cosmology (which is the subject of the  present work) as well as in the  Maxwell $U(1)$ gauge theory.

The basic idea   goes as follows. The fundamentally new type of energy advocated in the present work can be, in principle, studied in a tabletop experiment by measuring some specific corrections to the Casimir vacuum energy  in the Maxwell theory as suggested in   \cite{Cao:2013na,Zhitnitsky:2015fpa,Cao:2015uza,Yao:2016bps,Cao:2017ocv}.
This fundamentally new contribution to the Casimir pressure emerges as a result of tunnelling processes, rather than due to the conventional fluctuations of the propagating photons with two physical transverse polarizations. Therefore, it was coined as the Topological Casimir Effect (TCE).
The extra energy computed in \cite{Cao:2013na,Zhitnitsky:2015fpa,Cao:2015uza,Yao:2016bps,Cao:2017ocv} is the direct analog of the QCD non-dispersive  vacuum  energy   (\ref{vacuum_energy}), (\ref{vacuum_energy1})    which is the key player of the present work as it explicitly enters (\ref{Delta}), (\ref{consistency}), (\ref{consistency1}) in the main text.
In fact, an extra contribution to the Casimir pressure emerges in this system as a result of nontrivial holonomy similar to (\ref{polyakov}) for the Maxwell field. The nontrivial holonomy in $E\&M$ system  is  enforced by  the nontrivial boundary conditions imposed
 in refs \cite{Cao:2013na,Zhitnitsky:2015fpa,Cao:2015uza,Yao:2016bps,Cao:2017ocv}, and related to the  nontrivial mapping $\pi_1[U(1)]=\mathbb{Z}$
 relevant for  the Maxwell abelian gauge theory. Furthermore, the ``reheating epoch" when the physical particles can be emitted from the vacuum in a time-dependent background, similar to the dynamical Casimir Effect, can be also tested in the Maxwell system as argued in \cite{Yao:2016bps}.

 A similar  new type of energy can be, in principle, also studied in superfluid He-II system which also shows a number of striking similarities
 with non-abelian QCD as argued in \cite{Zhitnitsky:2016hdz}. For the  superfluid He-II system the crucial role plays the vortices which are classified by $\pi_1[U(1)]=\mathbb{Z}$ similar to the   abelian quantum fluxes studied in  the Maxwell system in \cite{Cao:2013na,Zhitnitsky:2015fpa,Cao:2015uza,Yao:2016bps,Cao:2017ocv}.

\subsection{\label{no-boundary} Cosmological density matrix vs no-boundary and tunneling states }
  We conclude this section with few comments on  status of the density matrix initial conditions in cosmology
(which is the key element of the present work) as compared to the well known no-boundary \cite{noboundary} and tunneling \cite{tunnel,Vachaspati-Vilenkin,discord} proposals for the wavefunction of the Universe.

  As is known, observer independent treatment of the no-boundary state leads to an insufficient amount of inflation. Phenomenologically, the volume weighting \cite{volume-weighting,Hawking:2007vf} or top-down approach \cite{replica} to the no-boundary state seem to resolve this issue but remain with the problem of consistency of complex tunneling geometries and normalizability of the quantum ensemble in cosmology.

  On the other hand, tunneling state has a rather uncertain ground based on the hyperbolic rather than Schroedinger nature of the Wheeler-DeWitt equation. No-boundary wavefunction within the Euclidean path integral construction represents a special quasi-vacuum state. The tunnelling state within the approach of path integration over Lorentzian geometries leads to non-normalizable wavefunction with unstable quantum matter and gravity perturbations.
 This fact has been known since \cite{Vachaspati-Vilenkin},  long  before the recent works \cite{Feldbrugge:2017kzv,Feldbrugge:2017fcc,Feldbrugge:2017mbc} which extended this criticism also to the no-boundary wave-function.

Diversity of the definitions of the no-boundary and tunneling states (defined either as propagators or solutions of the homogeneous Wheeler-DeWitt equation either in Euclidean or Lorentzian spacetime) as  discussed in \cite{Feldbrugge:2017kzv,Feldbrugge:2017fcc,Feldbrugge:2017mbc}  actually indicates that neither of these states have a rigorous canonical quantization ground. However, the critical verdict of  \cite{Feldbrugge:2017kzv,Feldbrugge:2017fcc,Feldbrugge:2017mbc}  invalidating both the no-boundary and tunneling states, though it requires deeper consideration, does not actually achieve its goal. This is because what is actually required  is not the construction of the wavefunction itself, but rather scattering amplitudes, mean values and probabilities generated by it. The step from the wavefunction (or the density matrix) to these quantities is very nontrivial and requires additional integration over the end points of the path integral histories.  This integration can also run along the complex contours of the steepest decent approximation, it can bear UV divergences and might lead to the effects invalidating the main conclusions of  \cite{Feldbrugge:2017kzv,Feldbrugge:2017fcc,Feldbrugge:2017mbc}\footnote{This extra integration will require the selection of saddle points in the complex plane via the technique similar to that of  \cite{Feldbrugge:2017kzv,Feldbrugge:2017fcc,Feldbrugge:2017mbc}  which can unpredictably alter the results of this work.}.

This is exactly what is done in the microcanonical density matrix setup of \cite{Barvinsky:2006uh,Barvinsky:2006tu,why} -- we do not calculate the density matrix itself, but directly go over to its partition function dominated by the real valued periodic history in Euclidean spacetime. The starting point is the microcanonical density matrix of a spaially closed cosmology, which is defined as a projector on the space of solutions of the Wheeler-DeWitt equations -- quantum Dirac constraints of the canonical quantization of gravity in physical Lorentzian spacetime \cite{why}. The periodicity of the relevant saddle-point histories directly follows from the tracing procedure for the normalization of the density matrix (see Fig.1), and their Euclideanization is the corollary of the fact that periodic solutions exist only in the imaginary (Euclidean) time, which is equivalent to the integration over the complex contour of the lapse ADM function \cite{why}.

Thus, our approach differs from the methods of \cite{noboundary,tunnel} and  \cite{Feldbrugge:2017kzv,Feldbrugge:2017fcc,Feldbrugge:2017mbc}  in two major points -- firstly, the microcanonical density matrix prescription for the initial state of the Universe rather than the pure state wavefunction and, secondly, the calculation of the physical quantity -- partition function -- rather than the wavefunction or the density matrix. Conceptual rigidity of this construction avoids ambiguities of the approach of \cite{noboundary,tunnel,Feldbrugge:2017kzv,Feldbrugge:2017fcc,Feldbrugge:2017mbc} and unambiguously leads to $\mathbb{S}^1$-compatification of the Euclidean time bearing the holonomy of the gauge field -- the corner stone of the strongly coupled nonperturbative QCD-like theory and its effect of generating the vacuum energy.

\section*{Acknowledgements}
This project was initiated  during the workshop ``Quantum Vacuum and Gravitation: Testing General Relativity in Cosmology", Mainz Institute for Theoretical Physics, Mainz, March 2017.
We  are thankful to the organizers of this  workshop  for creating an amazing atmosphere  and the   opportunity to discuss some unorthodox ideas during the official talks and special evening sessions.
The work of AB was supported by the RFBR grant No.17-02-00651 and by the Yukawa Institute for Theoretical Physics. AZ was supported in part by the Natural Sciences and Engineering Research Council of Canada.

\appendix
 \section{\label{review}The nature  of   the ``non-dispersive"  vacuum energy. }
The main goal of this  Appendix is to review a number of  crucial  elements relevant for our studies of the ``non-dispersive"  vacuum energy and its cosmological significance. First,
we start in subsection \ref{contact} with   explanation of  a highly nontrivial nature of  this type of the   vacuum energy in the Euclidean space time.

This type of the vacuum energy is well known to the QCD practitioners, while it is much less known in the GR and  cosmology  communities.
We think this ignorance  can be explained by  the fact this unusual  type of the  vacuum energy cannot be formulated in terms of conventional local propagating degrees of freedom. Precisely such a ``local" formulation  is  a conventional framework  for the cosmology community
when the inflation or the dark energy is described in terms of a scalar field, such as inflaton  $\Phi (x)$ with  specifically adjusted local potential $V[\Phi]$.  On the other hand,   this unusual type of energy   has been known to  the QCD community for quite some time. Furthermore,  this unusual  ``non-dispersive"  nature of the vacuum energy has been  supported  by    a numerous lattice simulations, see \ref{contact} with references and the details.

  We continue in section \ref{sec:holonomy}  by clarifying  the crucial role of the holonomy (\ref{polyakov}) in generating such type of energy. We review   few   known analytical calculations of this type of energy by   emphasizing  the role of the non-local holonomy (computed along $ \mathbb{S}^1$) which generates this unusual energy.
  The  $ \mathbb{S}^1$ in these  computations represents  an important  portion of a larger  Euclidean 4d manifold     $\mathbb{S}^3\times \mathbb{S}^1$, which has been extensively employed   in the main text of this work, see sections  \ref{Inflation} and \ref{Inflation-2}.

\exclude{
  Finally, in section \ref{interpretation} we list the main features of this novel type of energy relevant for the cosmology. Specifically, we elaborate  on non-local
  and non-analytical (with respect to the coupling constant)  properties of these  configurations saturating the ``non-dispersive" contribution to the   vacuum  energy. The corresponding analysis  plays an important role in the main   text in sections  \ref{Inflation} and \ref{Inflation-2}  where  we study the Euclidean gravitational  instantons  and argue that precisely this type of the vacuum energy saturates the partition function and generates the de Sitter behaviour   after the Lorentzian Universe nucleates from the Euclidean instanton.
}
  In Section \ref{inst-quark} we make few historical remarks on fractionally charged topological objects as they  intimately related
  to non-trivial holonomy defined on $\mathbb{S}^1$.

\subsection{\label{contact}The topological susceptibility and contact term in flat space-time}
We start our  short   overview on  the ``non-dispersive"  nature of the vacuum energy  by reviewing a  naively unrelated topic--  the formulation and resolution of the so-called $U(1)_A$ problem in strongly coupled QCD~\cite{witten,ven,vendiv}. We
   introduce  the topological susceptibility $\chi$ which is ultimately related to the vacuum energy $E_{\mathrm{vac}}(\theta=0)$
   as follows\footnote{We use the Euclidean notations  where  path integral computations are normally performed.}
\be
\label{chi}
 \chi =    \left. \frac{\partial^2E_{\mathrm{vac}}(\theta)}{\partial \theta^2} \right|_{\theta=0}= \lim_{k\rightarrow 0} \int \!\dd^4x e^{ikx} \la T\{q(x), q(0)\}\ra  ~~~~~
 \ee
where     $\theta$ parameter   enters the  Lagrangian   along with  topological density operator $q (x)=
\frac{1}{16 \pi^{2}} \mathrm{tr}[ F_{\mu\nu} \tilde{F}^{\mu\nu}]$ and $E_{\mathrm{vac}}(\theta)$ is the   vacuum energy density
computed for  the Euclidean infinitely large flat space-time. This $\theta$- dependent portion of the vacuum energy (computed  at $ \theta=0$)   has a number of unusual properties as we review below.
The corresponding properties  are  easier  to explain in terms of the correlation function (\ref{chi}), rather than in terms of the vacuum energy $E_{\mathrm{vac}}(\theta=0)$ itself.  The relation between the two  is given by eq. (\ref{chi}).

 Few comments are in order.
First of all,  the topological susceptibility $\chi$  does not vanish in spite of the fact that $q(x)= \partial_{\mu}K^{\mu}(x)$ is total derivative. This feature is very different from any conventional correlation functions  which normally must  vanish  at zero momentum $ \lim_{k\rightarrow 0}$  if the  corresponding operator  can be represented as  total divergence.

Secondly, any physical $|n\ra$   state gives a {\it negative} contribution to this
diagonal correlation function
\be	\label{G}
  \chi_{\rm dispersive} \sim  \lim_{k\rightarrow 0} \int d^4x e^{ikx} \la T\{q(x), q(0)\}\ra \nonumber \\
  \sim
    \lim_{k\rightarrow 0}  \sum_n \frac{\la  0 |q|n\ra \la n| q| 0\ra }{-k^2-m_n^2}\simeq -\sum_n\frac{|c_n|^2}{m_n^2} \leq 0,
\ee
 where   $m_n$ is the mass of a physical $|n\ra$ state,  $k\rightarrow 0$  is  its momentum, and $\la 0| q| n\ra= c_n$ is its coupling to topological density operator $q (x)$.
 At the same time the resolution of the $U(1)_A$ problem requires a positive sign for the topological susceptibility (\ref{chi}), see the original reference~\cite{vendiv} for a thorough discussion,
\be	\label{top1}
  \chi_{\rm non-dispersive}= \lim_{k\rightarrow 0} \int \!\dd^4x e^{ikx} \la T\{q(x), q(0)\}\ra > 0.~~~~~
\ee
Therefore, there must be a contact contribution to $\chi$, which is not related to any propagating  physical degrees of freedom,  and it must have the ``wrong" sign. The ``wrong" sign in this paper implies a sign
  which is opposite to any contributions related to the  physical propagating degrees of freedom (\ref{G}).
  The corresponding vacuum energy associated with non-dispersive contribution to the  topological susceptibility $\chi$ as defined by (\ref{chi})
  can be coined as  ``non-dispersive"  vacuum energy $E_{\mathrm{vac}}(\theta=0)$. It is quite obvious that the nature of   this energy  is drastically different from any types of conventional energy because it   {\it cannot} be formulated  in terms of any conventional propagating degrees of freedom  according to  (\ref{G}), (\ref{top1}). In the cosmological context relevant for the present work this type of energy in refs. \cite{Zhitnitsky:2013pna,Zhitnitsky:2014aja,Zhitnitsky:2015dia} was dubbed as the ``strange energy", while a scientific name would be  the ``non-dispersive"  vacuum energy $E_{\mathrm{vac}}(\theta=0)$
  generated by  the contact term in the correlation function (\ref{top1}). It should be contrasted with the ``dispersive" energy which, by definition, is   associated  with some   propagating degree of freedom  and  can be always restored from the absorptive portion of the correlation function through the dispersion relations according to (\ref{G}).

    In the framework \cite{witten} the contact term with ``wrong" sign  has been simply postulated, while in refs.\cite{ven,vendiv} the Veneziano ghost (with a ``wrong" kinetic term) had been introduced into the theory to saturate the required property (\ref{top1}).

  Our next comment is the observation that  the  contact term (\ref{top1}) has  the structure $\chi \sim \int d^4x \delta^4 (x)$.
  The significance of this structure is  that the gauge variant correlation function in momentum space
  \be
  \label{K}
   \lim_{k\rightarrow 0} \int d^4x e^{ikx} \la K_{\mu}(x) , K_{\nu}(0)\ra\sim   \frac{k_{\mu}k_{\nu}}{k^4}
   \ee
  develops  a topologically protected  ``unphysical" pole which does not correspond to any propagating massless degrees of freedom, but nevertheless must be present in the system. Furthermore, the residue of this   pole has the ``wrong sign" which    saturates the non-dispersive term  in gauge invariant correlation function (\ref{top1}),
   \be
  \label{K1}
   \< q({x}) q({0}) \> \sim  \la \partial_{\mu}K^{\mu}(x) , \partial_{\nu}K^{\nu}(0)\ra \sim \delta^4(x).
   \ee
 We conclude this review-type   subsection with the following remark. The entire framework, including the singular behaviour of
  $ \< q({x}) q({0}) \>$   with the ``wrong sign",  has been well confirmed by numerous  lattice simulations in strong coupling regime, and it is accepted by the community as a standard resolution of the $U(1)_A$ problem. Furthermore, it has been argued long ago in ref.\cite{Luscher:1978rn}
  that the gauge theories may exhibit the ``secret long range forces" expressed in terms of the correlation function (\ref{K}) with topologically protected pole at $k=0$.

  Finally, in a weakly coupled gauge theory (the so-called ``deformed QCD" model \cite{Yaffe:2008}) where all computations  can be performed in  theoretically  controllable way
one can explicitly test every single element of this entire framework, including the topologically protected pole (\ref{K}), the contact term with ``wrong sign", etc,  see ref. \cite{Thomas:2011ee,Thomas:2012ib,Zhitnitsky:2013hs} for the details. In particular, one can explicitly see that the Veneziano ghost is in fact an auxiliary topological field which saturates the vacuum energy and the topological susceptibility $\chi$.  It does not violate unitarity, causality and any other fundamental principles  of a quantum field theory. What is more important for the present studies is that  one can explicitly see that the holonomy (\ref{polyakov})  plays a crucial role   in generating of the ``strange" vacuum energy defined in terms of the correlation function  (\ref{chi}).

While all these unusual features of the vacuum energy are well-known and well-supported by numerous lattice simulations in strongly coupled regime (see e.g. \cite{Zhitnitsky:2013hs} for a large number of references on original lattice results)  a precise quantitative   understanding of these properties  (on a level of analytical computational scheme) is still  lacking.  In next subsection we review some known results on this matter specifically emphasizing on role of the holonomy (\ref{polyakov}) in the analytical computations. Precisely a nontrivial holonomy  (\ref{polyakov})   plays a crucial  role in generating of the ``strange" vacuum energy as we shall argue in next subsection \ref{sec:holonomy}. This is the key technical element  which pinpoints the source of this novel type of energy   not    expressible  in terms of any  local operators as the holonomy is obviously a non-local object.

\subsection{\label{sec:holonomy}The  holonomy (\ref{polyakov}) and generation of the ``non-dispersive" vacuum  energy.
   }
Our goal here is
to argue that the holonomy plays a key role in generation of the ``non-dispersive" vacuum  energy  in the system.
We also want to compare the vacuum energy computed on
  different manifolds such as   $\mathbb{S}^1\times\mathbb{R}^3$ versus    $\mathbb{S}^1\times\mathbb{H}^3$  and  $\mathbb{S}^1\times\mathbb{S}^3$. Such studies
    play the crucial  role  in our   analysis in the main text in section \ref{sec:grav-instanton} devoted to  construction of the gravitation instanton formulated on $\mathbb{S}^1\times\mathbb{S}^3$.

 We start our analysis with $\mathbb{S}^1\times\mathbb{R}^3$ geometry. The key role in the  discussions will play the behaviour of holonomy $U(\mathbf{x})\equiv{\cal{P}}\exp\left(i\int_0^{{\cal{T}}} dx_4 A_4(x_4, \mathbf{x})\right)$ at spatial infinity,  the Polyakov line,
  \be
\label{polyakov}
L={\cal{P}}\exp\left(i\int_0^{{\cal{T}}} dx_4 A_4(x_4, |\mathbf{x}|\rightarrow\infty)\right).
\ee
The operator $\Tr L$ classifies the self-dual  solutions which may contribute to the path integral at finite temperature $T\equiv {\cal{T}}^{-1}$, including the low temperature limit $T\rightarrow 0$.
There is a well known generalization of the standard self-dual instantons to non-zero temperature, which corresponds to
 the description on $\mathbb{R}^3 \times \mathbb{S}^1$ geometry. This is so-called periodic instantons, or calorons\cite{Harrington:1978ve}  studied in details in \cite{Gross:1980br}. These  calorons   have trivial holonomy, which implies that the $\Tr L$   assumes values belonging to the group centre $\mathbb{Z}_N $ for the $SU(N)$ gauge group.

More general class of the self-dual  solutions with nontrivial holonomy (\ref{polyakov}), the so-called KvBLL calorons  were constructed much later in refs.  \cite{Kraan:1998pm,Lee:1998bb}. In this case the holonomy (\ref{polyakov}) in general,  is not reduced to the
group centre $\Tr L\notin \mathbb{Z}_N$. The fascinating feature of the  KvBLL calorons is that they can be viewed as a set of $N $ monopoles of $N$ different types.  Normally, one expects that monopoles come in $N-1$ different varieties carrying a unit magnetic charge from each of the $U(1)$ factors of the $U(1)^{N-1}$ gauge group left unbroken by vacuum expectation value due to nontrivial holonomy (\ref{polyakov}). There is an additional, so called  Kaluza- Klein (KK) monopole which carries magnetic charges and instanton charge. All monopole's charges are such that when complete set of  different types of monopoles are present, the magnetic charges exactly cancel, and the configuration of $N$ different monopoles carries a unit instanton charge.
In particular, for $SU(2)$ gauge group the  holonomy
\be
\label{holonomy1}
\frac{1}{2} \Tr  L =\cos (\pi\nu),
\ee
belongs to the group center  $\frac{1}{2} \Tr  L=\pm 1$  when $\nu$ assumes the integer values (trivial holonomy). The so-called ``confining" value for the holonomy   corresponds to  $\nu=1/2$ when  $\Tr  L=0$ vanishes.

It has been known since \cite{Gross:1980br}
that the gauge configurations with non-trivial holonomy are strongly suppressed in the partition function. Therefore, naively
KvBLL calorons can not produce a finite contribution to the partition function. However, this   naive argument is  based on consideration of the individual KvBLL caloron, or finite number of them. If one considers a  grand canonical  assemble of these objects than  their density is determined by the dynamics, and    the old argument  of ref. \cite{Gross:1980br}
does not hold anymore.  The corresponding objects in this case  may in fact   produce a finite contribution to the partition function.  A self consistent
computations in a weak coupling regime supporting this picture have been carried out in the so-called ``deformed QCD" model
\cite{Yaffe:2008}. One can explicitly see how $N$ different types of   monopoles with nontrivial holonomy (\ref{polyakov}) which carry fractional topological charge $\pm 1/N$ produce  confinement, generate the ``strange" vacuum  energy (\ref{chi}) and associated with this energy the topological susceptibility (\ref{K1})  with known, but highly unusual properties reviewed above in previous subsection \ref{contact}, see  \cite{Thomas:2011ee,Thomas:2012ib,Zhitnitsky:2013hs} for the technical details on these computations.

In the strong coupling regime we are interested in, the corresponding analytical computations  have never been completed.
There is a limited  number of partial analytical and numerical results  \cite{Diakonov:2004jn,Diakonov:2007nv,Liu:2015ufa} on computations of moduli space and one loop determinant,  controlling   the dynamics and interaction properties of the  constituents in a large  ensemble of KvBLL calorons.

While complete analytical solution in strong coupling regime is still lacking, nevertheless  there is a number of hints supporting the basic picture  that the KvBLL configurations with nontrivial holonomy (\ref{polyakov}) and representing $N$ different types of monopoles with fractional topological charges $\pm 1/N$ saturate the ``strange" vacuum  energy (\ref{chi}) and associated with this energy the topological susceptibility (\ref{K1}) in a very much the same way as it happens in
``deformed QCD" model  where all computations are performed in a theoretically controllable regime \cite{Yaffe:2008,Thomas:2011ee,Zhitnitsky:2013hs}.
It is assumed in what follows  that  the topological susceptibility (\ref{chi}) and associated   with it the  ``non-dispersive" vacuum  energy   $E_{\mathrm{vac}}(\theta)$  is  indeed saturated by  fractionally charged monopoles with $Q=\pm 1/N$ which are constituents of KvBLL caloron with nontrivial holonomy (\ref{polyakov}), (\ref{holonomy1}).

The corresponding computations of  the partition function and the free energy for the vacuum ground state  for $\mathbb{S}^1\times\mathbb{R}^3$ geometry lead to the following result
  \cite{Diakonov:2004jn,Diakonov:2007nv,Liu:2015ufa}:
 \be
 \label{Z3}
 { \cal{Z}}\simeq \exp\left[4\pi fV\right], ~~~~ f=\frac{4\pi\Lbar^4}{g^4 T} \nonumber\\ F_{\rm vac}=-T\ln  { \cal{Z}}=-\frac{32\pi^2}{g^4} \Lbar^4 V,
\ee
 where $V$ is the 3-volume of the system, $g$ is the coupling constant of a non-abelian gauge field, the    $\Lbar$
 is a single dimensional parameter of the system generated as a result of dimensional transmutation in classically conformal gauge theory, similar to conventional $\Lambda_{\rm QCD}\simeq 170$ MeV in
 QCD physics.     Parameter $f$ in (\ref{Z3})  can be interpreted as the monopole's fugacity of  the system, while the combination ${\cal{T}}F_{\rm vac} \equiv E_{\rm vac} V^{(4)}\equiv  E_{\rm vac} {\cal{T}} V$ shows the extensive property when $\ln  { \cal{Z}}$  is  proportional to the Euclidean 4-volume at  large $V^{(4)}\rightarrow \infty$. In this framework  $E_{\rm vac}$ has dimension $4$ and represents  the vacuum energy density of the system
   entering   the fundamental formula  (\ref{chi}) and  defining  the ``non-dispersive" portion of the vacuum energy.

  One can show that free energy   (\ref{Z3}) as well as the topological susceptibility  $\chi$ demonstrate   all the features of the ``strange energy" briefly described in section \ref{contact} in model-independent generic way, including the ``wrong" sign for $\chi$ which cannot be associated with any physical propagating degrees of freedom.
    The specific  mechanism based on the KvBLL configurations reviewed above  and  describing the tunnelling processes between the distinct topological sectors  precisely generates all these required properties. In what follows we assume that the very same mechanism generates the ``non-dispersive" vacuum energy density  $E_{\rm vac}$ for different geometries, including $\mathbb{S}^3 \times \mathbb{S}^1$ and $\mathbb{H}^3_{\kappa}\times \mathbb{S}^1_{\kappa^{-1}}$ exactly in the same way as computed  above for $\mathbb{R}^3 \times \mathbb{S}^1$.

    With this assumption  in hand the question we address below   is as follows.
How does the ``strange energy"  density $E_{\rm vac}$  vary if the geometry is slightly modified at   large distances?
The main motivation for this question is originated from our fundamental conjecture formulated in sections \ref{Inflation} that the energy density which enters the Friedman equation represents in fact the difference $\Delta E$ between the energy density computed in a nontrivial background by subtracting the ``trivial" portion computed in the flat background similar to the Casimir type computations.

Specifically,  we want to know how  does the vacuum energy density  depend on  the geometry $\mathbb{S}^3 \times \mathbb{S}^1$. Precisely this information is required   in our computations  in sections \ref{Inflation} and \ref{Inflation-2}.  Unfortunately, there is a number of technical obstacles  to carry out the computations similar   to (\ref{Z3}) for $\mathbb{S}^3 \times \mathbb{S}^1$ manifold. In particular, even the   monopole solution (which is the crucial ingredient in this type of semiclassical computations) satisfying the appropriate boundary conditions on $\mathbb{S}^3 \times \mathbb{S}^1$ is not exactly known.
As a result of this deficiency, a semi-classical computation which would account for zero and non-zero modes contributions to the partition function
(similar to formula (\ref{Z3}) derived for $\mathbb{R}^3 \times \mathbb{S}^1$) is also not known.

Fortunately, the exact semi-classical computations are  available for the hyperbolic space
$\mathbb{H}^3_{\kappa}\times \mathbb{S}^1_{\kappa^{-1}}$. While this manifold is not exactly what we need for  our analysis in sections  \ref{Inflation} and \ref{Inflation-2}, nevertheless, the  corresponding computations  give us a hint on possible corrections to  the vacuum energy density $E_{\rm vac}$ due to a small dimensional parameter  $\sim \kappa$ which emerges in the $\mathbb{H}^3_{\kappa}\times \mathbb{S}^1_{\kappa^{-1}}$ in comparison with computations (\ref{Z3}) corresponding to $\mathbb{R}^3 \times \mathbb{S}^1$ geometry.

The main reason why the semiclassical computations can be carried out in  hyperbolic space $ \mathbb{H}^3_{\kappa}$  with the constant negative curvature
$-\kappa^2 $ is as follows. There is a conformal equivalence between $(\mathbb{R}^4- \mathbb{R}^2) $ and  $ \mathbb{H}^3_{\kappa}\times \mathbb{S}^1_{\kappa^{-1}}$ where  $\mathbb{S}^1_{\kappa^{-1}}$ denotes the circle of radius $\kappa^{-1}$. As a result of this exact equivalence, the monopole's  solutions  can be explicitly constructed in this case. The holonomy (\ref{polyakov}) is computed   along a closed loop $\mathbb{S}^1_{\kappa^{-1}}$ and    assumes a nontrivial value.

 The key observation of this computation, see formula (\ref{ratio}) below,  is that  the topological configurations with non-trivial holonomy
 produce a finite contribution to the vacuum energy density  with a small correction being  linearly proportional to $\kappa\rightarrow 0$. This  effect can not be expressed in  terms of any local operators such as curvature as $|R|\sim \kappa^2$. Rather, the leading correction $\sim \kappa$ is generated due to topological  vacuum configurations with nontrivial holonomy, not expressible in terms of any  local observables. This is precisely the reason why the generic arguments \cite{Shapiro:1999zt,Shapiro:2000dz,Sola:2013gha} based on locality simply do not apply here.

Now we are ready to formulate the main result of the computations \cite{Zhitnitsky:2015dia} relating
   the vacuum energy density $E_{\rm vac}$ computed on  the original $\mathbb{R}^3 \times \mathbb{S}^1$ manifold and on the hyperbolic space $\mathbb{H}^3_{\kappa}\times \mathbb{S}^1_{\kappa^{-1}}$.
   In formula (\ref{ratio}) presented below  we assume that  the sizes of $\mathbb{S}^1$ from two different manifolds are identically the same, i.e. we identify
 ${\cal{T}}=\kappa^{-1}$. After this identification the only difference between two manifolds   is the curvature of the hyperbolic space $R[\mathbb{H}^3_{\kappa}]\sim \kappa^2$ at $\kappa\rightarrow 0$.
Formula (\ref{ratio}) below suggests a linear dependence on $\kappa$ at small $\kappa$ which we interpret   as a strong argument supporting our conjecture   on linear dependence of ``non-dispersive" vacuum energy as a function of external parameter.  Such linear scaling obviously implies that this background-dependent correction is not related to  any local operators such as curvature, but rather is generated by  nonlocal operator   (\ref{polyakov})  which is sensitive to the global characteristics of the background.

The relevant formula can be represented as follows  \cite{Zhitnitsky:2015dia}:
\be
\label{ratio}
\frac{E_{\rm vac}[ \mathbb{H}^3_{\kappa}\times \mathbb{S}^1_{\kappa^{-1}}]}{E_{\rm vac}[ \mathbb{R}^3 \times \mathbb{S}^1]}
 \simeq  \left(1-  \frac{\nu(1-\nu)}{2}\cdot \frac{\kappa}{\Lbar} \right).
\ee
Using formula (\ref{Z3}) the same result can be written as follows
\be
\label{vacuum_energy}
&&E_{\rm vac}[ \mathbb{H}^3_{\kappa}\times \mathbb{S}^1_{\kappa^{-1}}]\simeq   -\frac{32\pi^2}{g^4} \Lbar^4 \left(1-  \frac{\nu(1-\nu)}{2}\cdot \frac{\kappa}{\Lbar} \right)\nonumber\\
&&\simeq -\frac{32\pi^2}{g^4} \Lbar^4 + \frac{32\pi^2}{g^4} \Lbar^3\cdot \frac{\nu(1-\nu)}{2}\cdot \kappa.
\ee
The key observation here is that a small correction here is linear, rather than naively expected quadratic function   at small $\kappa\rightarrow 0$.
Furthermore, the correction $\sim \kappa$ vanishes  for configurations with trivial holonomy, $\nu=0, \nu=1$.

This observation unambiguously implies that
 the relevant Euclidean configurations which are capable to produce the linear correction (\ref{vacuum_energy}) must carry a nontrivial holonomy (\ref{polyakov}), and therefore, they are non-local in nature. The   computations \cite{Thomas:2012ib} in weakly coupled ``deformed QCD" model  (where a configuration with nontrivial holonomy produces a linear correction) also support this claim.

\subsection{Generation of the Holonomy in a strongly coupled gauge theory}\label{inst-quark}
The question we want to address in this Appendix can be formulated as follows.
If we consider the thermodynamical limit in eq. (\ref{Z3}) one can explicitly see that
  the combination ${\cal{T}} F_{\rm vac} \equiv E_{\rm vac} V^{(4)}\equiv  E_{\rm vac} {\cal{T}} V$ shows the extensive property when $\ln  { \cal{Z}}$  is  proportional to the Euclidean 4-volume at  large $V^{(4)}\rightarrow \infty$. In this framework  $E_{\rm vac}$ has dimension $4$ and represents  the vacuum energy density of the system entering    the fundamental formula  (\ref{chi}). This formula   defines   the ``non-dispersive" $\theta$ dependent portion of the vacuum energy which plays the crucial role in our analysis.

  The key question we want to address now is as follows:   if we start from description of the system on $\mathbb{R}^4$ from the very beginning
 such that the semiclassical solutions (calorons with nontrivial holonomy) cannot be constructed on $\mathbb{R}^4$. How do we know  anything    about the holonomy defined on $\mathbb{S}^1$ (and its direct consequence in form of the objects with fractional topological charges)  if it was not a part of our construction to begin with? We should emphasize here that the configurations with fractional topological charges is very strong signal that there is a nontrivial holonomy in the system as the only semiclassical  solutions which can be defined on $\mathbb{R}^4$ are integer value instantons.

 We obviously do not know the answer on the hard  question formulated above in strongly coupled 4D QCD. However, there is a well known example of the 2D $CP^{N-1}$ model which hints that such kind of holonomy (and its manifestation in form of the configurations with fractional topological charges)  might be generated dynamically by strong quantum fluctuations such that the ``effective calorons" with nontrivial holonomy do appear in the system, but they are strongly coupled quantum objects, rather than the semiclassical configurations defined on $\mathbb{S}^1$.

 Historically, the configurations with fractional topological charges emerged in 2D $CP^{N-1}$ model. These fractional objects have been coined as
 instanton quarks, also known as ``fractional instantons" or ``instanton partons".
Namely, using an exact accounting and re-summation of the $n$-instanton
solutions in 2d CP$^{N-1}$ models, the original statistical problem of a
 grand canonical instanton  ensemble (with exclusively integer topological charges defined on $\mathbb{R}^2$) was mapped unto a 2d Coulomb Gas
system of pseudo-particles with fractional topological charges $\sim
1/N$ \cite{Fateev:1979dc,Berg:1979uq}.   This picture leads to the elegant explanation of
the confinement phase and other important properties of the
$2d~CP^{N-1}$ models \cite{Fateev:1979dc,Berg:1979uq}.   The   term  ``instanton quarks"  was introduced to  emphasize that there are precisely $N$ constituents making an integer instanton, similar to $N$ quarks making a baryon.   These objects do not appear individually in path integral; instead, they appear as configurations consisting $N$ different   objects with fractional charge $1/N$ such that the total topological charge of each configuration   is always  integer. In this case $2Nk$ zero modes for $k$ instanton solution is interpreted as $2$   translation zero modes modes accompanied by  every single instanton quark.  While the instanton quarks emerge in the path integral coherently,
  these objects are highly delocalized: they may emerge on opposite sides of the space time or be close to each other with alike   probabilities.
Similar attempt to   in 4D  QCD was unfortunately unsuccessful due to a number of technical problems, which remain to be solved \cite{Belavin:1979fb}.

There is deep   analogy with ``deformed QCD" model \cite{Yaffe:2008,Thomas:2011ee, Zhitnitsky:2013hs} where the size of  $\mathbb{S}^1$ is fixed   for the semiclassical approximation to be justified. However, it is a common  view in the QCD community that the physics in strongly coupled QCD is qualitatively the same as in weakly coupled ``deformed QCD" model with enforced semi-classicality by specifically chosen $\mathbb{S}^1$ in which case  the configurations with nontrivial holonomy (and fractionally charged monopoles) can be explicitly constructed  on the semiclassical level. Furthermore, it is expected that even
in a  case when the corresponding $\bar{\theta}$ parameter in strongly coupled $\qcd$ does not vanish, the physics remains the same and the confinement in $\qcd$ occurs as a result of condensation of the same fractionally charged monopoles as argues in \cite{Anber:2017rch}.

The main lesson to be learnt in the context of the present work is as follows. The configurations with fractional topological charges can serve as a trigger
for a nontrivial holonomy because conventional semiclassical solutions defined on $\mathbb{R}^4$ can carry only integer topological charges.  The lesson from  2d CP$^{N-1}$ is as follows. The fractional topological charges are not present in the system when it is defined on $\mathbb{R}^2$.
However, such objects do appear dynamically as a result of strong quantum fluctuations.  In terms of effective semiclassical configurations these objects obviously require a nontrivial holonomy (and therefore, nontrivial   $\mathbb{S}^1$ where the holonomy is defined). However, this effective $\mathbb{S}^1$
is not the original circle, but rather the effective one which emerges as a result of strong quantum fluctuations. This is precisely the motivation for our
Model-2 in section \ref{Inflation-2} where we unlink the size of $\mathbb{S}^1$ from matter context of the theory by relaxing the bootstrap equation.
Unfortunately, we can only speculate on this matter at present time without making any precise and solid claims.

\section{Topological auxiliary field as a non-propagating and non-dynamical   inflaton} \label{BF-section}
The goal of this Appendix is to introduce the auxiliary field technique and  demonstrate that the corresponding alternative computations reproduce the crucial elements  of the vacuum energy and its unusual features   listed in section \ref{interpretation}.
Furthermore, this technique will play a crucial role in our studies on anomalous coupling with the SM fields described   in section \ref{anomaly}. Precisely this coupling is responsible for the successful {\it reheating} phase as advocated in  sections \ref{instability}, \ref{e-folding}.

As we  argue below we  can identify (on intuitive level) the corresponding auxiliary non-dynamical, non-propagating field with
the {\it inflaton}, which is an emergent field in our framework: it only appears  in the confined $\qcd$ phase, while in the  deconfined phase it did not exist in the system.  It should be contrasted with conventional description in terms of local dynamical field $\Phi(x)$ which always a part of  the  system, long before and long after the inflation.

 We should emphasize that the reformulation of the same physics in terms of an auxiliary  quantum  field rather than in terms of explicit
computation of the partition function by summing over all topological sectors   is not a mandatory procedure, but a matter of convenience. Similarly, the description of a topologically ordered phase in condensed matter physics in terms of Chern Simons effective Lagrangian is a matter of convenience rather than a necessity as emphasized in Section \ref{anomaly}.

We shall  demonstrate how this technique works in a simplified version of QCD, the so-called  weakly coupled ``deformed QCD'' model \cite{Yaffe:2008} which preserves all relevant features of the strongly coupled QCD such as confinement, nontrivial $\theta $ dependence, generation of the ``non-dispersive" vacuum energy, etc. At the same time, all computations can be performed under complete theoretical control.    The   computations  of the ``non-dispersive''  term by explicit summation over positions and orientations of the monopoles-instantons describing the tunnelling transitions   have  been performed in \cite{Thomas:2011ee}. The corresponding results have been reproduced
in \cite{Zhitnitsky:2013hs} using the technique of the auxiliary  topological fields.

One should also mention that the computations are performed in effectively 3d  weakly coupled gauge theory, rather than in strongly coupled 4d $\qcd$.
Nevertheless, the emergent auxiliary field to be introduced below and identified with {\it inflaton} behaves, in all respects  as the 4d Veneziano ghost  \cite{ven,vendiv} which was postulated long ago precisely with the purpose to describe these unusual features of the vacuum energy as reviewed  in Appendix \ref{contact}.

 The basic idea to describe the relevant IR physics in terms of an auxiliary field is to insert the corresponding $\delta$- function into the path integral with a Lagrange multiplier and integrate out the fast degrees of freedom while keeping the slow degrees of freedom which are precisely the  auxiliary fields. Here and in what follows we use notations from \cite{Zhitnitsky:2013hs} where this technique was originally implemented to demonstrate that the famous Veneziano ghost is nothing but auxiliary topological field. The   $\delta$- function to be inserted into the path integral is defined as follows
 \be
\label{delta}
\delta \left( q(\mathbf{x})+ \frac{1 }{4 \pi NL}   \left[\vec{\nabla}^2 a (\mathbf{x}) \right]\right)\sim \nonumber\\
\int {\cal{D}}[b]e^{ i\int d^4x ~b(\mathbf{x})\cdot\left(q(\mathbf{x})+  \frac{1 }{4 \pi NL}   \left[\vec{\nabla}^2 a (\mathbf{x}) \right]\right)},
      \ee
     where  $q(\mathbf{x})\sim \mathrm{tr}
		\left[ F_{\mu\nu} \tilde{F}^{\mu\nu} \right] $ in this formula is treated as the original expression   for the topological density operator
		including the fast non-abelian gluon degrees of freedom, while $b(\mathbf{x}), a(\mathbf{x})$ are  treated as  slow-varying   external sources describing the large distance physics  for a given monopole configuration. One can proceed now with conventional semiclassical computations by summation over all monopoles, their positions and orientations  to arrive to the following dual form for the effective action. The new additional  topological term $\sim  b(\mathbf{x})\vec{\nabla}^2 a (\mathbf{x})$ can be immediately
		recovered from (\ref{delta}), while interaction of the $b(\mathbf{x})$ field (playing the role of the  Lagrange multiplier) coupled to topological density operator $q(\mathbf{x})$ can be easily recovered as it has   precisely the structure of  the $\theta$ term.
		This observation unambiguously implies that $b(\mathbf{x})$ field enters the effective description in unique combination with $\theta$ as follows $[\theta-b(\mathbf{x})]$ as long as $b(\mathbf{x})$ field can be treated as a slow degree of freedom.
		In all respects it is similar to construction of the effective Lagrangian for the $\eta'$ field which enters the action in
		unique combination with $\theta$ as follows $[\theta-\eta']$. The difference is, of curse, that $\eta'$ meson has a kinetic term as well, in contrast with $b(\mathbf{x})$ field.
		Therefore, the final expression for the dual effective  action which includes new auxiliary $b(\mathbf{x}), a(\mathbf{x}) $ fields assumes the form  \cite{Zhitnitsky:2013hs}:
				 \be
\label{b-action}
{\cal Z} [\bm{\sigma},   b, a]&\sim&\int {\cal{D}}[b]{\cal{D}}[\bm{\sigma}]{\cal{D}}[ a]e^{-S_{\rm top}[b, a]-S_{\rm dual}[\bm{\sigma},   b]},  \nonumber\\
S_{\rm top}[b, a] &=&
 \frac{-i }{4 \pi N}  \int_{\mathbb{R}^{3}}  d^{3}x     b(\mathbf{x})\vec{\nabla}^2 a (\mathbf{x}) ,\\
	S_{\rm dual}[\bm{\sigma},   b] &=& \int_{\mathbb{R}^{3}}  d^{3}x \frac{1}{2 L} \left( \frac{g}{2 \pi} \right)^{2}
		\left( \nabla \bm{\sigma} \right)^{2} \nonumber  \nonumber\\
		&-&\zeta  \int_{\mathbb{R}^{3}}  d^{3}x \sum_{a = 1}^{N} \cos \left( \alpha_{a} \cdot \bm{\sigma}
		+ \frac{\theta-b(\mathbf{x})}{N} \right).
		 \nonumber
\ee
  In this formula parameter $\zeta$ plays the role of the monopole's density in the system, such that the vacuum energy
  is explicitly proportional to $\zeta$, see (\ref{YM_top}) below. The dynamical $\bm{\sigma}$ fields effectively  describe
  the monopole's ensemble. The most important element for our studies is the  Lagrange multiplier
 $b(\mathbf{x})$ field and topological $a (\mathbf{x})$ field which  will be interpreted as the inflaton   in what follows.
 Both fields are not dynamical, and not propagating degrees of freedom, by construction. We obviously do not introduce any new dynamical degrees of freedom by inserting the $\delta$ function (\ref{delta}) and introducing the auxiliary topological fields $b(\mathbf{x}), a(\mathbf{x}) $. This is obviously important remark when one tries to identify $b(\mathbf{x}), a(\mathbf{x}) $ with inflaton.

   One next step is to compute the vacuum energy and topological susceptibility within this framework to demonstrate that it satisfies all the features listed in section \ref{interpretation}.
   The corresponding computations explicitly show that the  physical meaning of the vacuum energy is the number of the tunnelling events per unit volume per unit time
   The corresponding formula can be represented  in terms of  the   correlation function   as follows
    \be
\label{YM_top}
E_{\rm vac}&=& -N^2\lim_{k\rightarrow 0} \int d^4x e^{ikx} \la q(\mathbf{x}), q(\mathbf{0})\ra \\
&=&-\frac{N\zeta}{L}\int d^3x \delta^3(\mathbf{x})= -\frac{N\zeta}{L},   \nonumber
\ee
where we represented $q(\mathbf{x})$ in terms of the auxiliary field $ -\frac{1 }{4 \pi NL}   \left[\vec{\nabla}^2 a (\mathbf{x}) \right]$
and performed the Gaussian path integral over ${\cal{D}}[b]{\cal{D}}[\bm{\sigma}]{\cal{D}}[ a]$ fields entering (\ref{b-action}).

We obviously reproduce our previous result based on explicit computations of the monopoles \cite{Thomas:2011ee}. Now it is formulated in terms of the long -ranged auxiliary topological fields.   The fluctuating  $b(\mathbf{x}), a(\mathbf{x}) $ fields simply reflect the long distance dynamics of the degenerate topological sectors which exist independently from our description in terms of  $b(\mathbf{x}), a(\mathbf{x}) $ fields. However, in previous computations   \cite{Thomas:2011ee} we had to sum over all monopoles, their positions, interactions  and orientations. Now this problem is simplified as it is reduced to the computation of the correlation function constructed from the auxiliary fields governed by the action (\ref{b-action}).

We identify (intuitively) the corresponding auxiliary
$[a(\mathbf{x}), b(\mathbf{x})] $ fields  which saturate this energy (\ref{YM_top}) with {\it inflaton} in this model
in  a sense that both objects eventually lead to the deSitter behaviour.
  We emphasize again that the corresponding dynamics can not be formulated in terms of a
 canonical scalar field $\Phi$ with any local potential $V(\Phi)$ as it is known that the   dynamics governed by CS-like action is truly non-local.   There is a large number of CM systems (realized in nature)  where CS action plays a key role with explicit manifestation of the  non-locality in the system.   It has been also argued that the deformed QCD model which is explored in this section also belongs to a topologically ordered phase   with many  features which normally accompany the topological phases~\cite{Zhitnitsky:2013hs}.

 What is the physical meaning of this auxiliary   $[a(\mathbf{x}), b(\mathbf{x})] $ fields  which we identify with {\it inflaton}?
 What is the best way  to  visualize it  on the intuitive level?
 From our construction one can easily see that  both fields  $[a(\mathbf{x}), b(\mathbf{x})] $  do not carry a colour index. However,  $a(\mathbf{x})$ field has nontrivial  transformation
 properties under large gauge transformation. In fact
 our field $ \nabla_i a(\mathbf{x}) $ transforms  as the $K_{i} (\mathbf{x})$ in the Veneziano construction (\ref{K}).    One can support this identification  by computing  a gauge variant  correlation function
 \be
 \label{K_top}
 \lim_{k\rightarrow 0} \int d^4x e^{ikx} \la \nabla_i a(\mathbf{x}), \nabla_j a(\mathbf{0})\ra \sim\frac{k_i k_j}{k^4}.
 \ee
  The  massless  pole (\ref{K_top}) has precisely the same nature as the pole  in the Veneziano construction  (\ref{K}).

 What is the physical meaning of   $b(\mathbf{x}) $ field? This field can be thought as an external axion $\theta(x)$ field, without kinetic term, though.

  Our comment here is that in spite of the gap $\sim \zeta$ in the system, some correlation functions constructed from the topological auxiliary fields $a(\mathbf{x}), b(\mathbf{x})  $ fields are still highly sensitive to the IR physics.   Furthermore, while the behaviour (\ref{K_top}) at small $k$ can be considered to be very dangerous as it includes $k^4$ in denominator (which normally attributed to the negative norm states in QFT), the physics described here is perfectly unitary and
 causal as $a(\mathbf{x}), b(\mathbf{x})  $ are in fact auxiliary rather than propagating dynamical fields as all questions can be formulated and answered even without mentioning the auxiliary topological fields.

 One should comment here that the results presented above are based on computations  in a weakly coupled, effectively 3d, gauge theory (where the system is   under complete theoretical control), while we are interested in 4d strongly coupled $\qcd$ to study the inflationary phase. Nevertheless,  the relation between the $a(x)$ auxiliary field
   and 4d $K_{\mu}$ field still holds  and  assumes the form
 \be
 K_{\mu}\sim \partial_{\mu}a(x),~~~~ q(x)\sim \partial_{\mu}K^{\mu}\sim \Box a(x)
  \ee
 while $b(x)$ field always enters the effective Lagrangian  precisely in combination with the $\theta$ term according to (\ref{b-action}) This observation  allows us to exactly reconstruct
 the interaction with SM particles  from the knowledge on their coupling to the $\theta$ parameter as eq.(\ref{coup}) states.

 What are the typical fluctuation scales of the auxiliary quantum $a(x)$ and $b(x)$ fields? The answer is quite obvious: the typical fluctuations are of order $\Lbar$ as the UV fluctuations of order $M_P$ are present  in the original gluon fields, but not in the auxiliary  $a(x)$ and $b(x)$ fields which effectively describe the  long distance physics in eq. (\ref{b-action}) where fast degrees of freedom are integrated  out.

 What happens when the same system is defined on a nontrivial manifold characterized by some dimensional parameters such as ${\cal{T}}^{-1}\ll \Lbar$  ? In this case  the fields $a(x, H)$ and $b(x, H)$ will continue to fluctuate with typical frequencies $\Lbar$. However, the relevant correlation functions
  should demonstrate
the emergence of the linear corrections with respect to these small parameters  $\sim {\cal{T}}^{-1}$. In particular, the correlation functions such as (\ref{YM_top}) computed in terms of the auxiliary fields are order of $\Lbar^4$   with corrections of order $(\Lbar{\cal{T}})^{-1}$  in agreement with expression (\ref{ratio1}) in section \ref{interpretation}.

We also want to make few comments on a typical scale of the expectation value of the field itself $\la b(x,H)\ra$ because, for example,  $\la \dot{b}(x,H)\ra$ enters the estimate (\ref{e-folds}) for the $e$-folds. As we discussed in the previous paragraph, the typical expectation values for the auxiliary fields must be expressed in terms of $\Lbar$ according to their dimensionality   because it reflects the typical topological density distribution in strongly coupled $\qcd$. According to our general prescription formulated in sections \ref{holonomy}, \ref{interpretation} we must subtract all the expectation values   computed on $\mathbb{R}^4$corresponding to $H=0$. This procedure unambiguously implies\footnote{For this specific  case
the particle production obviously does not occur when  $H=0$ as all topological transitions  simply select a specific $\theta$ sector, but do not generate the particle production. Therefore, the subtraction in this case is  trivial.}  that $\la \dot{b}(x,H)\ra\sim H$ as it must vanish at $H=0$. One should comment here is that $b(x,H)$ is not a classical field which satisfies some equations, similar to conventional studies on inflation when the potential $V[\Phi]$ completely determines the dynamics of the system. In our case one has to compute the relevant correlation functions and the expectation values to answer the questions about observables. We   make few additional comments on relation with conventional approach in section \ref{relation}.

 We want to present one  additional argument  supporting the same claim on linear correction $\sim {\cal{T}}^{-1}$ or $\sim H$.  The behaviour (\ref{K_top})   hints on possibility of  non-local effects (which indeed are known to be present in this system  \cite{Zhitnitsky:2013hs}). Such IR sensitivity suggests that the physics must be highly sensitive
 to the properties of the manifolds and external background configurations. Precisely this sensitivity to large distances supports our analysis    of section \ref{interpretation} where we argued that the corrections to the vacuum energy due to the finite manifold should be linear ${\cal{T}}^{-1}$ rather than exponential as a conventional  gapped theory would naively suggest.

Now we can   infer the physical meaning of the auxiliary fields: $a(x,H)$ describes the longitudinal portion of $K_{\mu}$ field generating the topologically protected pole ({\ref{K_top});  the $\Box a(x,H)$ describes the distribution of the topological density in the system; finally, $b(x,H)$ acts as the axion field (without kinetic term) being the source of the topological density distribution.

\end{document}